\documentclass[conference]{IEEEtran}
\IEEEoverridecommandlockouts
\usepackage{cite}
\usepackage{amsmath,amssymb,amsfonts}
\usepackage{graphicx}
\usepackage{textcomp}
\usepackage{xcolor}
\usepackage{algpseudocode}

\usepackage{hyperref}

\usepackage[linesnumbered, ruled, vlined]{algorithm2e}
\usepackage{amsmath,amsthm,amssymb}
\renewcommand{\algorithmicrequire}{ \textbf{Input: }} 
\renewcommand{\algorithmicensure}{ \textbf{Output: }} 
\usepackage[linesnumbered, ruled, vlined]{algorithm2e}
\usepackage{subfig}
\usepackage{balance}
\usepackage{booktabs} 
\newcommand{\ZYW}[1]{\textcolor{red}{#1}}

\newcommand{\del}[1]{}
\def\BibTeX{{\rm B\kern-.05em{\sc i\kern-.025em b}\kern-.08em
    T\kern-.1667em\lower.7ex\hbox{E}\kern-.125emX}}
\begin{document}
\makeatletter
    \newcommand{\linebreakand}{
      \end{@IEEEauthorhalign}
      \hfill\mbox{}\par
      \mbox{}\hfill\begin{@IEEEauthorhalign}
    }
    \makeatother

\title{Preserving Data Privacy in Learning Causal Structure with Fully Homomorphic Encryption
}

 \author{Jian Yang$^{1,2}$, Tong Yuan$^{1}$, Qinbin Li$^{3}$, Zeyi Wen$^{1,2}$, Xiaofang Zhou$^{2}$\del{\thanks{$^*$Corresponding author.}} \\
$^{1}$Hong Kong University of Science and Technology (Guangzhou),\\
$^{2}$Hong Kong University of Science and Technology,
$^{3}$University of California, Berkeley\\
\{jyang827,tyuan053\}@connect.hkust-gz.edu.cn, qinbin@berkeley.edu, \{wenzeyi, zxf\}@ust.hk 
}
\maketitle
\thispagestyle{plain}
\pagestyle{plain}
\begin{abstract}
Preserving data privacy is an important topic in structural data management and data mining. However, the issue of privacy leakage in distributed causal structure learning is a persistent challenge, especially in cases where data transmission and computation are required. In this paper, we propose a method based on fully homomorphic encryption (FHE) that performs calculations on ciphertexts, keeping data encrypted in transition and computation. Nevertheless, adopting FHE to causal structure learning is challenging due to the high computation cost and limited support on division as well as logarithm operations in FHE. To tackle this challenge, we propose a series of novel techniques including (i) circuit simplification for better efficiency, (ii) approximation of division and logarithm through Newton-Raphson Reciprocal and Taylor expansion, and (iii) a batching technique with SIMD-acceleration to enhance the whole learning process. Additionally, our method can be easily extended beyond FHE by demonstration of its portability to support differential privacy. Empirical results show that our method achieves high consistency and comparable causal structure with the plaintext version in the datasets tested. Last, our method is efficient and practical to complete learning causal structures in tens of minutes even under the privacy protection of FHE.
\end{abstract}


\section{Introduction}
\del{Causal learning~\cite{pearl1988bn} aims to learn probabilistic graphical models that employ directed acyclic graphs (DAGs) to compactly represent a set of random variables and their conditional dependencies. 
The graphical nature makes them well-suited for representing knowledge with uncertainty and effective reasoning. 
With the recent growing demand for interpretable machine learning models, causal learning has attracted much research attention since they are inherently interpretable models~\cite{gunning2019darpa, rudin2019stop}.

The inherent graphical nature makes them particularly appropriate for representing knowledge pervaded with uncertainty and fostering efficient reasoning. 
}

Causal learning~\cite{pearl1988bn} aims to learn probabilistic graphical models that employ directed acyclic graphs (DAGs) to compactly represent a set of random variables and their conditional dependencies. The learned structures are widely applied in data management of the smart city, such as traffic prediction~\cite{zhan2016citywide} and automatic drive~\cite{chen2018rear}, thanks to its innate interpretability~\cite{gunning2019darpa, rudin2019stop}. Table~\ref{tab:exp} illustrates a dataset about traffic prediction and Figure~\ref{fig_bn_exp} shows a causal structure for this traffic prediction example. Each column in the table represents a variable and opposites a node in the causal structure, while each row is a sample that records the values of these variables. In most samples, when \textit{Heavy Rain} is true, \textit{Bad Vision} and \textit{Slippery Road} are true too, indicating \textit{Heavy Rain} may lead to \textit{Bad Vision} and \textit{Slippery Road}. Hence, two edges \textit{``Heavy Rain''-``Bad Vision''} and \textit{``Heavy Rain''-``Slipped Road''} exist among these three variables. The relation between other variables can be inferred from the data, such as \textit{Bad Vision} and \textit{Slipped Road} are reasons for \textit{Accident}. In addition, \textit{Accident} may lead to \textit{Congestion} and \textit{Injury}.

\begin{table}[ht]
\caption{Dataset of the causal structure example.}
    \footnotesize	
    \begin{center}
    \begin{tabular}{cccccc}
        \toprule
        Heavy Rain & Bad Vision & Slippery & Accident  & Congestion & Injury\\ 
        \midrule
        \textit{True} & \textit{False} & \textit{True} & \textit{True} & \textit{True} & \textit{True} \\
        \textit{True} & \textit{True} & \textit{True} & \textit{False} & \textit{True} & \textit{False} \\
        \textit{False} &  \textit{False} & \textit{True} & \textit{True} & \textit{True} & \textit{True} \\
        \textit{True}  & \textit{True} & \textit{True} &  \textit{False} & \textit{True} & \textit{False} \\
        \multicolumn{6}{c}{...} \\
        \bottomrule
    \end{tabular}
    \label{tab:exp}
    \end{center}
\end{table}

\begin{figure}[ht]
    \centering
    \includegraphics[width=0.8\linewidth]{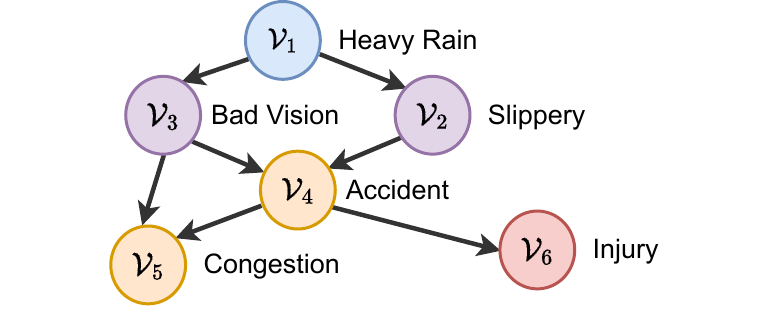}
    \caption{An example of causal structure.}
    \label{fig_bn_exp}
\end{figure}

The data-driven approach has been used for learning causal structure quantitatively and automatically, to encapsulate causal relationships among the variables. A cornerstone of structure learning algorithms is the PC-stable~\cite{colombo2014order}, which commences with a complete undirected graph and eliminates edges at sequential depths based on CI tests. Notably, the PC-stable algorithm has found widespread use in diverse applications~\cite{zhang2012inferring, maathuis2010predicting} and has been implemented in mainstream causal learning packages such as bnlearn~\cite{scutari2009learning} and tetrad~\cite{ramsey2018tetrad}. On the other hand, privacy-preserving causal structure learning represents a significant area of exploration. At present, fully homomorphic encryption (FHE)~\cite{gentry2009fully}, capable of directly executing addition and multiplication operations on ciphertexts, addresses issues related to data transmission and computation. This makes it particularly fitting for privacy-focused causal structure learning. However, the following three major challenges are associated with the learning: (i) the computation circuit of FHE for causal structure learning is complex and large, requiring careful design to guarantee correctness and efficiency; (ii) division and $\log$ operations in causal structure learning are hard problems in FHE~\cite{aslett2015review}; (iii) the complex circuit for structure learning further aggravates the time-consuming issue of FHE.

To overcome these challenges in the FHE-based method, we propose a privacy-preserving causal structure learning system to handle this secure computation problem.
Firstly, we reduce the complexity of computation circuits and the number of computation operations by substituting fractions and terms within the equations. 
Secondly, we carefully design arithmetic circuits to approximate division and logarithm by addition and multiplication, while ensuring the retainment of efficiency and precision. 
Thirdly, we develop several techniques for enhancing efficiency and precision, which include (i) the SIMD techniques to compress numerous conditionally independent (CI) tests into one batch and expedite execution, thereby significantly enhancing the speed of the proposed method; (ii) deliberated initial guesses and expansion to reduce the number of iterations and boost the performance; (iii) optimizing the communication cost by transferring only crucial data.
Lastly, we demonstrate the portability of our method by adapting it to differential privacy. To sum up, we make the following major contributions in this paper.
\begin{itemize}
    \item We propose a privacy-preserving causal structure learning method in distributed systems by fully homomorphic encryption, which isolates the plaintext of computation and the executor of computation and guarantees plaintext security. The proposed method is versatile and can be effortlessly extended to accommodate differential privacy.
    \item To support division and logarithm, we design two approximate circuits to compute the division by Newton-Raphson reciprocal and logarithm by Taylor expansion, ensuring precision and efficiency.
    \item Our proposed method is equipped with a series of techniques for good efficiency, including batching techniques to enhance efficiency by grouping many CI tests into a single batch for concurrent execution, better initial guesses for reciprocal and expansion points for logarithm to converge faster, and transferring only crucial data to save communication overhead.
    \item We conduct extensive experiments to validate the correctness, efficiency, and scalability of our method. The results prove that our method can complete the task efficiently and correctly, and can be extended across multiple machines. 
\end{itemize}

\del{The paper is organised as follows: In Section~\ref{paper:rw}, we reviewd the corelated work in causal structure learning and privacy-aware distributed learning, in Section~\ref{sec_pre}, we claimed all the preliminaries that are used in this paper, in Section \ref{sec_alg}, we described the design of the arithmetic circuit we used to preform FHE based structure learning and how we enhance its efficiency, and Secition~\ref{sec_exp}, show the experient result of proposed method including accuracy, preformance rescourse requirement as well as how the hyperparameters in the algoritm are carefully selected to ensure optimal performance.}

\section{Preliminaries}
\label{sec_pre}
\del{In this section, we first provide the key terminologies and definitions related to causal structure learning and review the learning algorithms. 
Then we introduce two privacy-protection methods: differential privacy (DP) and fully homomorphic encryption (FHE). Differential privacy protects personal information by adding noise in individual data drawn from a zero-mean distribution while fully homomorphic encryption performs addition and multiplication operations on ciphertext directly without decryption.}

In this section, we initially provide the terminologies and definitions related to causal structure learning and review the learning algorithms. Then, we present two methodologies for safeguarding privacy: differential privacy (DP) and fully homomorphic encryption (FHE). DP preserves personal information by infusing noise into individual data that does not change the statistics of interest, while FHE enables the execution of addition and multiplication operations directly on ciphertext without decryption.

\subsection{Causal Structures}

Causal structures are a class of probabilistic graphical models that represent a joint distribution $P$ over a set of random variables $\mathcal{V} = \{\mathcal{V}_1, \mathcal{V}_2, ... , \mathcal{V}_{n}\}$ via a DAG. Typically, one variable corresponds to one feature in  machine learning problems.
We use $\mathcal{G} = (\mathcal{V}, \mathcal{E})$ to denote the DAG. In a DAG $\mathcal{G}$, each node in $\mathcal{V}$ is associated with one variable, and each edge in $\mathcal{E}$ represents conditional dependencies among the two variables. $\mathcal{V}_j$ is called a parent of $\mathcal{V}_i$ if there exists a directed edge from $\mathcal{V}_j$ to $\mathcal{V}_i$ in $\mathcal{G}$, and we use $Pa(\mathcal{V}_i)$ to denote the set of parent variables of $\mathcal{V}_i$.

In a causal structure, each variable has its local probability distribution 
that describes the probabilities of possible values of this variable given its possible parent configurations. The joint probability of variables $\mathcal{V}$ in a causal structure can be decomposed into the product of local probability distributions of each variable, and each local probability distribution only depends on its correlated variable $\mathcal{V}_i$ and its parents: 
\\$P(\mathcal{V}_1, \mathcal{V}_2, \dots,\mathcal{V}_{n}) = \prod_{i=1}^{n} P(\mathcal{V}_i | Pa(\mathcal{V}_i))$,
where $n$ is the number of variables in $\mathcal{V}$, $P(\mathcal{V}_1, \mathcal{V}_2, ...,\mathcal{V}_{n})$ represents the joint probability of the variables and $P(\mathcal{V}_i | Pa(\mathcal{V}_i))$ represents the conditional probability of variable $\mathcal{V}_i$.

\subsection{Conditional Independence Tests}

Consider some random variables $\mathcal{V}_i$, $\mathcal{V}_j$ and $\mathcal{V}_k$ in a causal structure, a CI test assertion of the form $I(\mathcal{V}_i, \mathcal{V}_j | \{\mathcal{V}_k\})$ means $\mathcal{V}_i$ and $\mathcal{V}_j$ are independent given $\mathcal{V}_k$. Given dataset $\mathcal{D}$, a CI test $I(\mathcal{V}_i, \mathcal{V}_j | \{\mathcal{V}_k\})$ determines whether the corresponding hypothesis $I(\mathcal{V}_i, \mathcal{V}_j| \{\mathcal{V}_k\})$ holds or not. In the domain of discrete variables, two commonly employed primary statistical tests are \del{For discrete variables, the two most common statistics for testing $I(\mathcal{V}_i, \mathcal{V}_j| \{\mathcal{V}_k\})$ are} the $\chi^2$ and $G^2$ tests~\cite{spirtes2000causation} defined as:
\begin{equation}
    \label{equ_chi2}
    \chi^2 = \sum_{x, y, z} \frac{(N_{xyz} - E_{xyz})^2}{E_{xyz}}
\end{equation}
\begin{equation} 
\label{equ_g2}
G^2 = 2 \sum_{x, y, z} N_{xyz} \log \frac{N_{xyz}}{E_{xyz}},
\end{equation}
where $N_{xyz}$ is the number of samples in $\mathcal{D}$ that satisfy $\mathcal{V}_i = x$, $\mathcal{V}_j = y$ and $\mathcal{V}_k = z$. The value of $N_{xyz}$ can be obtained from the contingency table that shows the frequencies for all configurations of values. 
$G^2$ could be considered as a simple version of $\chi^2$ and follows an asymptotic $\chi^2$ distribution with $(|\mathcal{V}_i|-1)(|\mathcal{V}_j|-1)$,
where $|\cdot|$ denotes the number of possible values of the variable. The p-value of $\chi^2$ distribution can be calculated according to the $G^2$ statistic and the final decision is made by comparing the p-value with the significance level $\alpha$. If p-value is greater than $\alpha$, the independent hypothesis $I(\mathcal{V}_i, \mathcal{V}_j| \{\mathcal{V}_k\})$ is accepted; otherwise, the hypothesis is rejected.
$E_{xyz}$ is the expected frequency which is defined as
\begin{equation} 
\label{equ_e}
E_{xyz} = \frac{N_{x+z} N_{+yz}}{N_{++z}},
\end{equation}
where $N_{x+z} = \sum_{y} N_{xyz}$, $N_{+yz} = \sum_{x} N_{xyz}$, and $N_{++z} = \sum_{xy} N_{xyz}$, which represent the marginal frequencies.

\subsection{The Causal Structure Learning Algorithm}
The PC-stable algorithm~\cite{colombo2012learning} is one of the most popular methods for causal structure learning from data. PC-stable consists of three steps. The first step is to determine the skeleton of the graph. The term skeleton means the underlying undirected graph of the learned structure. This step is done by performing a large number of CI tests. The second step is to identify the v-structures in the skeleton. A v-structure is a triple $(\mathcal{V}_i, \mathcal{V}_j, \mathcal{V}_k)$ that can be denoted by $\mathcal{V}_i \rightarrow \mathcal{V}_k \leftarrow \mathcal{V}_j$. In other words, nodes $\mathcal{V}_i$ and $\mathcal{V}_j$ have an outgoing edge to node $\mathcal{V}_k$ and are not connected by any edge in the graph. 
The v-structure is a key component to distinguish different causal structures. By identifying the v-structures in this step, some edges in the skeleton become directed edges. The third step is to set directions for as many of the remaining undirected edges as possible by applying a set of rules called Meek rules~\cite{meek2013causal}. For example, we set the direction of the undirected edge $\mathcal{V}_j - \mathcal{V}_k$ into $\mathcal{V}_j \rightarrow \mathcal{V}_k$ whenever there is a directed edge $\mathcal{V}_i \rightarrow \mathcal{V}_j$ such that $\mathcal{V}_i$ and $\mathcal{V}_k$ are not adjacent; otherwise a new v-structure is created. In the three steps of the PC-stable algorithm, only the first step involves data exchange~\cite{zarebavani2019cupc}, and hence we focus on privacy protection for the first step, which is further elaborated to protect the privacy of the learning process in Section~\ref{sec_alg}.

\begin{algorithm} [tb]
\DontPrintSemicolon
\LinesNumbered
\caption{Causal Structure Learning Algorithm.}   
\label{alg_pc_stable}

\algorithmicrequire Node set $\mathcal{V}$, Dataset $\mathcal{D}$

\algorithmicensure  Graph $\mathcal{G}$, $SepSet$

Graph $\mathcal{G}\xleftarrow{}$  FormCompletedGraph($\mathcal{V}$) $= \mathcal{V} \times \mathcal{V}$

Depth $d \xleftarrow{} 0$

Let $a(\mathcal{V}_i)$ represent adjacency nodes of $\mathcal{V}_i$ 

\textbf{repeat} 

\quad \textbf{for} any edge $(\mathcal{V}_i, \mathcal{V}_j)$ in $\mathcal{G}$ \textbf{do} 
 
\quad \quad \textbf{repeat}

\quad \quad \quad Choose a new $\mathcal{S} \subseteq a(\mathcal{V}_i)\backslash\{\mathcal{V}_j\}$ with $|\mathcal{S}| = d$ 

\quad \quad \quad Perform CI test $I(\mathcal{V}_i, \mathcal{V}_j | \mathcal{S})$

\quad \quad \quad \textbf{if} hypothesis $I(\mathcal{V}_i, \mathcal{V}_j | \mathcal{S})$ holds \textbf{then}

\quad \quad \quad \quad Remove $(\mathcal{V}_i, \mathcal{V}_j)$ from $\mathcal{G}$

\quad \quad \quad \quad Store $\mathcal{S}$ in $SepSet(\mathcal{V}_i, \mathcal{V}_j)$

\quad \quad \quad \textbf{end if}

\quad \quad \textbf{until} $(\mathcal{V}_i, \mathcal{V}_j)$ is removed or all $\mathcal{S}$ are considered

\quad $d \xleftarrow{} d + 1$

\textbf{until} all pairs of $(\mathcal{V}_i, \mathcal{V}_j)$ in $\mathcal{G}$ satisfy $|a(\mathcal{V}_i)\backslash\{\mathcal{V}_j\}| < d$
 
\end{algorithm}   

The pseudo-code of the key part of the causal structure learning with PC-stable is given in Algorithm~\ref{alg_pc_stable}. 
The general idea is to initialize $\mathcal{G}$ to a complete undirected graph over the node set $\mathcal{V}$ (Line 3), and remove some of the edges determined as independent by performing a number of CI tests in consecutive depths (Lines 6 to 16).
Specifically, at each depth $d$, the algorithm iteratively records the current adjacency sets of all the nodes, where $a(\mathcal{G}, \mathcal{V}_i)$ denotes the adjacent nodes of $\mathcal{V}_i$ in $\mathcal{G}$ (Line 5). This operation is used for choosing the separating set $\mathcal{S}$ later.
Next, for every edge $(\mathcal{V}_i, \mathcal{V}_j)$ in the graph $\mathcal{G}$, a number of CI tests $I(\mathcal{V}_i, \mathcal{V}_j | \mathcal{S})$ are performed for different separating sets. The elements in the separating sets are chosen from $a(\mathcal{V}_i)\backslash\{\mathcal{V}_j\}$, and the size of each separating set $|\mathcal{S}|$ is equal to the current depth $d$ (Lines 7 to 10). If there exists a separating set $\mathcal{S}$ where $\mathcal{V}_i$ is independent of $\mathcal{V}_j$ given $\mathcal{S}$, the edge $(\mathcal{V}_i, \mathcal{V}_j)$ is removed from $\mathcal{G}$, and $\mathcal{S}$ is stored in $SepSet(\mathcal{V}_i, \mathcal{V}_j)$ (Lines 11 to 13). $SepSet(\mathcal{V}_i, \mathcal{V}_j)$ denotes the separating set of $\mathcal{V}_i$ and $\mathcal{V}_j$, which is used in the second step of the PC-stable algorithm to identify the v-structures. Since the second step is fast and is not the focus of our work, we omit the details of separating set. Once all edges are considered, $d$ is incremented (Line 15) and the above procedure is repeated for the next depth. Depth $d$ is used to control the size of the separating sets from small to large. 
This process continues until all pairs of adjacent nodes $(\mathcal{V}_i, \mathcal{V}_j)$ in $\mathcal{G}$ satisfy $|a(\mathcal{V}_i)\backslash\{\mathcal{V}_j\}| < d$ as shown in Line 16. 

\subsection{Differential Privacy}

Differential privacy (DP) provides a quantifiable privacy assurance that prevents the exposure of individual's sensitive information in a dataset, while allowing useful computations on the aggregate data \cite{dwork2006differential}. 
The objective of DP ensures that the inclusion or exclusion of a single individual's data does not significantly affect the total results.
Formally, a randomized algorithm $\mathcal{M}$ satisfies $\epsilon$-DP if for any two adjacent datasets $\mathcal{D}$ and $\mathcal{D}'$ that differ in at most one record, and any subset $S$ of the range of $\mathcal{M}$, the probability of $\mathcal{M}(\mathcal{D})$ outputting $S$ is bounded by:
\begin{equation*}
    Pr[\mathcal{M}(\mathcal{D}) \in S] \leq e^{\epsilon} \cdot Pr[\mathcal{M}(\mathcal{D}') \in S]
\end{equation*}
where $\epsilon > 0$ is a parameter that controls the strength of privacy guarantee and $\delta$ is a small constant that represents the probability of the algorithm deviating.

The goal of DP in distributed machine learning is to ensure that the learning process does not leak the privacy of personal data~\cite{shokri2015privacy}. This is usually achieved by adding carefully calibrated random noises to the computation process which prevents reconstruction of individual records from the shared outputs.
To ensure consistency with the origin algorithm, the added noise should balance the privacy and utility~\cite{abadi2016deep}. This balance is crucial because if too much noise is added, the utility of the data can be compromised while adding too little noise can risk privacy.

There are various techniques to add noise like the Laplace Mechanism and Exponential Mechanism~\cite{huang2012exponential, dwork2006differential}. Each of these approaches varies in their theoretical privacy guarantees, computational requirements, and impact on the final model’s accuracy. 
DP provides robust privacy guarantees in the face of side-channel attacks and post-processing. Hence, it has gained considerable interest in the research community for its potential to protect privacy in distributed machine learning systems.

\subsection{Fully Homomorphic Encryption}
Fully homomorphic encryption (FHE)~\cite{gentry2009fully} has become a cornerstone in the field of privacy-preserving machine learning. It allows computations to be performed on encrypted data without requiring access to the decryption key, thereby ensuring data privacy 
FHE operations mainly consist of two types: addition and multiplication. Let $c_1$ and $c_2$ be two ciphertexts encrypting plaintexts $m_1$ and $m_2$, respectively. The homomorphic properties of FHE can be defined as follows:

\begin{itemize}
\item \textbf{Addition:} $Dec_{sk}(c_1 + c_2) = m_1 + m_2$
\item \textbf{Multiplication:} $Dec_{sk}(c_1 \cdot c_2) = m_1 \cdot m_2$
\end{itemize}

where $Dec_{sk}$ denotes the decryption function using secret key $sk$.
The Brakerski-Fan-Vercauteren (BFV)~\cite{fan2012somewhat} and Boneh-Goh-Nissim (BGV)~\cite{brakerski2014leveled} schemes support computations on integers and binary data. 
The Cheon-Kim-Kim-Song (CKKS) scheme is a variant of FHE that supports approximate arithmetic on complex numbers, making it suitable for handling floating-point data \cite{cheon2017homomorphic}. The scale $\Delta$ in CKKS is a parameter that controls the precision of the computations. The encryption of a message $m$ in CKKS can be represented as $c = Enc_{pk}(m; \Delta)$, where $pk$ is the public key. The decryption of a ciphertext $c$ is $Dec_{sk}(c) \approx m$, where the approximation is due to the inherent noise in the CKKS scheme.

In the CKKS scheme, batching is a technique in FHE that allows a single ciphertext to encrypt a vector of plaintexts, thereby enabling single instruction multiple data (SIMD) operations. This is particularly useful in machine learning applications where computations are often performed on large datasets. In the context of FHE, SIMD operations can be performed in parallel on the encrypted data, leading to significant speedup.

\section{Method}
\label{sec_alg}

This section introduces our proposed privacy-preserving causal structure learning method in the distributed setting. The overview of our proposed method is shown in Figure~\ref{fig_alg_overall}. According to the figure, the coordinator starts the process by raising a CI test and notifying all the machines. Then, each machine computes local results, encrypts the local results, and sends ciphertexts to the coordinator. Later, the coordinator computes global results according to ciphertexts of local results, obtains encrypted global results, and sends global results back to machines. Finally, each machine decrypts global results, and updates the local network referring to global results. The coordinator aggregates local results and produces global results. 

Our proposed method is applicable to typical distributing settings. In a typical distributed learning scenario, the training dataset is stored in multiple machines. Each worker machine stores a distinct partition of the training dataset. A typical training dataset comprises several samples, each containing multiple dimensions. In the context of causal structure learning, these dimensions are referred to as nodes. Each sample consists of the values of these nodes, while each node can have varied values across different samples. Unlike the single machine scenario where each process can access the entire dataset and distribute tasks among nodes, here, processes are restricted to access only a subset of the training data. These processes execute tasks in parallel on their specific samples, following the instructions set forth by the coordinator.
\begin{figure}
    \centering
    \includegraphics[width=\linewidth]{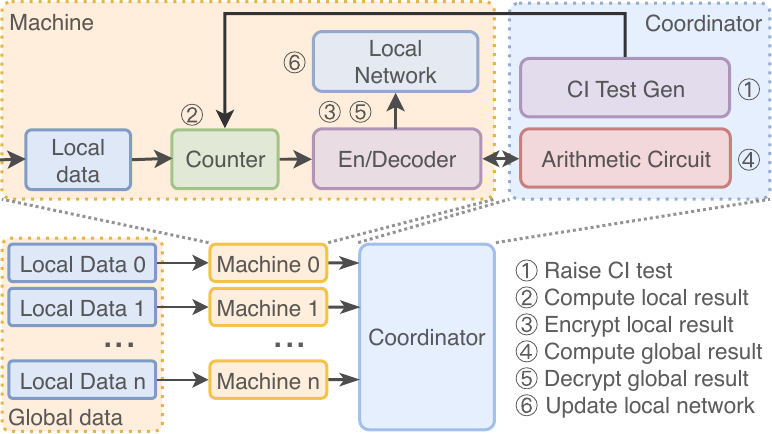}
    \caption{The pipeline of our proposed privacy-preserving distributed causal structure learning method.}
    \label{fig_alg_overall}
\end{figure}

In the following of this section, the technical details of our proposed method are introduced in four parts. 
\begin{itemize}
    \item First, we describe distributed causal structure learning where each machine holds a subset of the training data. Each machine computes the local results based on the subset it holds, and sends local results to the coordinator. This method enables the distributed causal structure learning method on horizontally partitioned data to output identical learning results as the causal structure method on the entire data.
    \item Second, we present the threat model of our privacy-preserving distributed causal structure learning. There are two threats from the computation process due to coordinators' curiosity and data leakage in communication channels. The coordinator may infer some sensitive information from machines' data collected in computation. Communication channels transfer correct information but leak information to attackers. We show that these two threats can be solved by privacy-preserving techniques, such as homomorphic encryption.
    \item Third, we elaborate on the details of our proposed full privacy protection solution with fully homomorphic encryption (FHE). FHE supports applying addition and multiplication directly on ciphertexts without decryption. With this property, we design a conditional independent (CI) test system where machines encrypt their individual local results and transmit encrypted local results to the coordinator. The coordinator performs arithmetic operations on encrypted local results and obtains encrypted global results by FHE in complete unawareness of the plaintext. For $\log (x)$ and $1/x$ functions in CI tests, we design arithmetic circuits to calculate them by FHE addition and multiplication. In addition, we accelerate the execution of CI tests by the SIMD-style batching technique. 
    \item Finally, to prove the portability to other privacy-preserving algorithms, we illustrate the differential privacy version of our proposed privacy-preserving causal structure learning. The mechanism builds a noise distribution whose mean is zero and draws a noise from it for each data sum operation. The noise avoids reverse reasoning that discriminates individual data, while zero-mean offsets the errors caused by noise.
\end{itemize}

\subsection{Causal Structure Learning in Plaintext}
\label{paper:plaintext}
To decide the existence of an edge in the causal structure, we conduct CI tests based on separating sets from neighbor nodes of its two endpoints.
The depth is equal to the size of the separating set, increasing one by one at each level. The size of the separating set affects the computation efficiency: smaller separating sets have lower computation costs when testing the existence of edges, while the cost grows exponentially with the increase of the size~\cite{van2010fully}. 
When performing the CI test, the coordinator machine raises a CI test request containing an edge and a separating set. It notifies worker machines to calculate partial results based on the subset they hold. After that, worker machines send their partial results to the coordinator to obtain the final results, and then the worker machines decide whether to delete the edge according to the final results.

Considering the CI tests in Equations~\eqref{equ_chi2} and~\eqref{equ_g2}, calculating the value of $G^2$ and $\chi^2$ need to compute $E_{xyz}$ and $N_{xyz}$ first. In a single machine setting, it is easy to count $E_{xyz}$ and $N_{xyz}$ in one pass. 
However, in multiple machine settings, each machine needs to count local results and the coordinator merges local results into one global result.
The term $N_{xyz}$ can be computed by aggregating the partial/local results $N_{xyz}^i$ from each worker machine, i.e., $N_{xyz} = \sum^i N_{xyz}^i$. Similarly, other terms can be computed using the following formulas: $N_{x+z} = \sum^{i} N_{x+z}^i$,
$N_{+yz} = \sum^{i} N_{+yz}^i$,
$N_{++z} = \sum^{i} N_{++z}^i$,
where $N_{xyz}^i$, $N_{x+z}^i$, $N_{+yz}^i$, and $N_{++z}^i$ represent the local $N_{xyz}$, $N_{x+z}$, $N_{+yz}$, and $N_{++z}$ of the $i$-th worker machine, respectively. For each CI test, we need to aggregate these values to obtain global results that consider all the disjoint subsets of the entire training dataset. 

The pseudo-code of the causal structure learning algorithm in a distributed environment is provided in Algorithm~\ref{alg_disjoint_pc_stable}.
We add a machine set $\mathcal{M}$ and training dataset $\mathcal{D}$ as the input (Line 1) for this setting.
The initial part keeps the same as the Algorithm~\ref{alg_pc_stable} (Lines 3-5).
For the computation part (Lines 6-19), we first raise a new CI test request and inform the request to the worker machines (Line 8).
Then each machine $\mathcal{M}_i \in \mathcal{M}$ computes local results referring to their local dataset (Line 10) and sends local results to the coordinator $\mathcal{M}_0$ (Line 11).
Following that, the coordinator $\mathcal{M}_0$ collects local data and calculates global results (Line 12).
$\mathcal{M}_0$ compares global results with the p-value (Line 13) and determines the existence of the edge (Lines 14-16).
After finishing the CI test in this level $d$, the program turns to the next level to conduct CI tests (Line 18).

\begin{algorithm} [tb]
\DontPrintSemicolon
\LinesNumbered
\caption{Causal Structure Learning in Plaintext.}   
\label{alg_disjoint_pc_stable}

\algorithmicrequire Node set $\mathcal{V}$, Machine set $\mathcal{M}$, Dataset $\mathcal{D}_i \in  \mathcal{D}$, 
\text{\quad \quad  } $\mathcal{M}_i$ holds $\mathcal{D}_i$,  $\mathcal{M}_0$ is Coordinator

\algorithmicensure  Graph $\mathcal{G}$, $SepSet$

Graph $\mathcal{G}\xleftarrow{}$  FormCompletedGraph($\mathcal{V}$) $= \mathcal{V} \times \mathcal{V}$

Depth $d \xleftarrow{} 0$

Let $a(\mathcal{V}_i)$ represent adjacency nodes of $\mathcal{V}_i$ 

\textbf{repeat}

\quad \textbf{repeat}

\quad  \quad $C$ raise a CI test $I(\mathcal{V}_i, \mathcal{V}_j|\mathcal{S})$

\quad \quad \textbf{parallel for} $each$ $machine$ $\mathcal{M}_i$  \textbf{do} 

\quad \quad \quad Count $N_{xyz}^i$, $N_{x+z}^i$, $N_{+yz}^i$, and $N_{++z}^i$ in $\mathcal{D}_i$

\quad \quad \quad Send these four values to $\mathcal{M}_0$

\quad \quad \textbf{end parallel for}

\quad \quad $\mathcal{M}_0$ calculate $N_{xyz}$, $N_{x+z}$, $N_{+yz}$, and $N_{++z}$ by local values

\quad \quad $\mathcal{M}_0$ obtain p-value and compare p-value with threshold

\quad \quad \textbf{if} hypothesis $I(\mathcal{V}_i, \mathcal{V}_j | \mathcal{S})$ holds \textbf{then}

\quad \quad \quad Remove $(\mathcal{V}_i, \mathcal{V}_j)$ from $\mathcal{G}$

\quad \quad \quad Store $\mathcal{S}$ in $SepSet(\mathcal{V}_i, \mathcal{V}_j)$

\quad \quad \textbf{end if }

\quad \textbf{until} $(\mathcal{V}_i, \mathcal{V}_j)$ is removed or all $\mathcal{S}$ are considered

\quad $d \xleftarrow{} d + 1$

\textbf{until} all pairs of $(\mathcal{V}_i, \mathcal{V}_j)$ in $\mathcal{G}$ satisfy $|a(\mathcal{V}_i)\backslash\{\mathcal{V}_j\}| < d$

\end{algorithm}   

\subsection{Threat Model}
In the context of our threat model, we make certain assumptions to delineate potential risks and vulnerabilities in our system. Specifically, we assume that all the machines involved in the learning process, as well as the coordinator, are honest actors who strictly adhere to the established protocol for conducting computation and communication. However, we acknowledge that the coordinator may harbor a certain level of curiosity, seeking to infer sensitive information from the data provided by the machines. Our primary objective is to ensure the confidentiality and privacy of the data while allowing the computation to proceed. To this end, we outline the key threats and countermeasures within our threat model.

\textbf{Threat 1: Coordinator's Curiosity:} The coordinator may attempt to infer sensitive information from the data provided by the machines during computation, compromising the privacy of the individuals or entities whose data is being processed.

\textit{Countermeasure:} To mitigate this threat, we must ensure that the data remains secret throughout the computation conducted on the coordinator. This can be achieved through the use of advanced cryptographic techniques, including homomorphic encryption. Homomorphic encryption allows computations to be performed on encrypted data, preserving privacy while still obtaining the desired results.

\textbf{Threat 2: Data Leakage through Communication Channels:} During the communication phase, communication channels may transfer correct data but inadvertently leak data to unauthorized parties, compromising the integrity of the data.

\textit{Countermeasure:} To protect against this threat, we must implement mechanisms that enable secure and private communication. This involves employing cryptographic techniques such as asymmetric cryptography to verify the correctness of data transmission without exposing sensitive information.

\subsection{Learning with Fully Homomorphic Encryption}
Here, we illustrate the fully homomorphic encryption version of the causal structure learning method. The fully homomorphic encryption supports multiplication and addition, which can be extended to various complex operations. Our proposed method is based on Cheon-Kim-Kim-Song (CKKS)~\cite{cheon2017homomorphic} scheme, supporting float-point addition and operations that are required by the CI tests. A nice property of CKKS is that it allows computing a batch of inputs in one pass, which can significantly enhance learning efficiency when used properly. 

\del{When learning the structure, we need to perform the CI tests, either based on $G^2$ or $\chi^2$. To calculate the value of $\chi^2$ and $G^2$, we need to compute division and $\log$ operation refer to Equations~\eqref{equ_chi2} and~\eqref{equ_g2}. Here we adopt Newton-Raphson reciprocal and Taylor expansion techniques to approximate the real values of the operations. Moreover, in fully homomorphic encryption, computing one CI test has a fixed time consumption and a fixed amount of memory cost determined by the complexity of the arithmetic circuit. Since the resource consumption for one CI test is small and fixed, a machine can support many CI tests simultaneously. Inspired by SIMD parallel techniques, we execute a batch of CI tests simultaneously to shorten the elapsed time. }

During the structure learning process, CI tests, either based on $G^2$ or $\chi^2$, are essential. To ascertain the value of $\chi^2$ and $G^2$, it is necessary to perform division and $\log$ operations, referring to Equations~\eqref{equ_chi2} and~\eqref{equ_g2}. For these operations, we employ Newton-Raphson reciprocal and Taylor expansion techniques to approximate the real values. Additionally, in the context of fully homomorphic encryption, a single CI test requires a fixed amount of time and memory resources, determined by the complexity of the arithmetic circuit. Given the modest and constant resource consumption for a single CI test, a machine can concurrently support multiple CI tests. Drawing inspiration from SIMD parallel techniques, we concurrently execute a batch of CI tests to minimize the total time required.

\subsubsection{Arithmetic Circuit for $1/x$}
In fully homomorphic encryption, each computation task has an arithmetic circuit, which is recognized as a task execution flow graph. 
Let us take the $G^2$ case as an example, we compute $N_{xyz}$ and $E_{xyz}$ first. Then we calculate $1/E_{xyz}$ by Newton-Raphson reciprocal formula.
After that, we multiply $1/E_{xyz}$ with $N_{xyz}$ and obtain $N_{xyz}/E_{xyz}$. In the Newton-Raphson reciprocal calculation, the formula is derived from Newton's method by considering the reciprocal function $f(x) = 1/x$ and solving for the root of $f(x) - r = 0$, where $r$ is the reciprocal estimate. The resulting iteration formula is:
\begin{equation}
     r_{t+1} = r_t(2 - xr_t)
     \label{eq_rcp}
\end{equation}
which converges quadratically to the true reciprocal value. The reciprocal computation executes iteratively for several times to obtain the precision results. Figure~\ref{fig_rcp_circuit} illustrates the procedure of the arithmetic circuit for $1/x$. We input $x$ and an initial guess, then iterate computing Equation~\eqref{eq_rcp} $t$ times, and obtain approximate $1/x$.

\begin{figure}[ht]
    \centering
\includegraphics[width=0.7\linewidth]{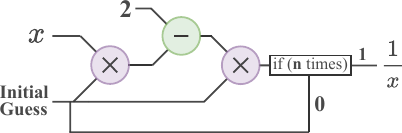}
     \caption{Newton-Raphson reciprocal circuit for $1/x$.}
    \label{fig_rcp_circuit}
\end{figure}

\subsubsection{Arithmetic Circuit for $\log(x)$}
The calculation of $G^2$ involves computing $\log\frac{N_{xyz}}{E_{xyz}}$ which can be approximated by Taylor expansion of $\log$ in $x_0 = 1$, because $N_{xyz}/E_{xyz}$ is around $1$. Once we finish these steps, we directly compute the product of $\log\frac{N_{Xyz}}{E_{xyz}}$ with $N_{xyz}$ and sum the product over every $xyz$ tuple. More specifically, the three-term Taylor expansion for $f(x)$ on $x_0 = a$ is as follows:
 \begin{equation*}
     f(x) \approx f(a) + f'(a)(x-a) + \frac{f''(a)}{2!}(x-a)^2.
 \end{equation*}
 Considering $N_{xyz}/E_{xyz}$ is near $1$ because $N_{xyz}$ is close to $E_{xyz}$ in most cases, we expand $\log$ on $x_0 = 1$:
 \begin{equation*}
     \log(x) \approx (x - 1) - \frac{(x - 1)^2}{2} + \frac{(x-1)^3}{3}.
 \end{equation*}
Here we implement the arithmetic circuit of the above equation using only multiplication and addition to approximate the $\log$ function. The three-term expansion has enough precision for the CI test, while more terms lead to higher computation costs but slight performance improvement, due to the exponential growth of the number of multiplication operations. Figure~\ref{fig_log_circuit} shows the three-term Taylor expansion circuit for approximate $\log(x)$. We compute $x^1$, $x^2$ and $x^3$, then multiply them with coefficients $1$, $\frac{1}{2}$ and $\frac{1}{3}$. Finally, we assemble by addition and subtraction together to obtain $\log(x)$. 

\begin{figure}[hb]
    \centering
    \includegraphics[width=0.8\linewidth]{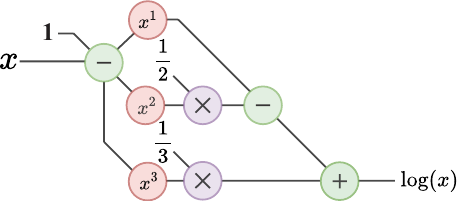}
    \caption{Taylor expansion circuit for $\log(x)$.}
    \label{fig_log_circuit}
\end{figure}

\subsubsection{Calculation for $\chi^2$ Test}
After building $1/x$ and $\log(x)$ circuits, we can implement $\chi^2$ and $G^2$ CI test circuits. To calculate the value of $\chi^2$, we substitute Equation~\eqref{equ_e} into Equation~\eqref{equ_chi2}. If we calculate the value $\chi^2$ first, we need to conduct reciprocal twice. First, we compute $1/N_{++z}$, and then we require to calculate $1/E_{xyz}$. The complexity of arithmetic grows up exponentially when the depth of the arithmetic circuit increases. 
So we reduce the fraction by multiplying $N_{++z}^2$ to both the denominator and numerator and obtain a new fraction that only needs one reciprocal.
The derivation is shown as follows:
\begin{equation*}
    \begin{aligned}
    \chi^2 &= \sum_{x, y, z} \frac{(N_{xyz} - E_{xyz})^2}{E_{xyz}} 
    = \sum_{x, y, z} \frac{(N_{xyz} - \frac{N_{x+z} N_{+yz}}{N_{++z}})^2}{\frac{N_{x+z} N_{+yz}}{N_{++z}}} \\
    &= \frac{(N_{xyz} N_{++z} - N_{x+z}N_{+yz})^2}{N_{+yz}N_{x+z}N_{++z}}
    \end{aligned}
\end{equation*}

Figure~\ref{fig_chi2_circuit} illustrates the arithmetic circuit of our proposed $\chi^2$ calculation method.
We obtain $N_{+yz}$, $N_{x+z}$, and $N_{++z}$ by marginalizing $N_{xyz}$ over $x$, $y$ and $z$, respectively. Then, we feed these values into the circuit. 
$(N_{xyz} N_{++z} - N_{x+z}N_{+yz})^2$ and $\frac{1}{N_{+yz}N_{x+z}N_{++z}}$ can be computed in parallel and incorporate them by a multiplication operation.
After the execution of the circuit, we obtain the $\chi^2$ value.
\begin{figure}[h]
    \centering
    \includegraphics[width=0.7\linewidth]{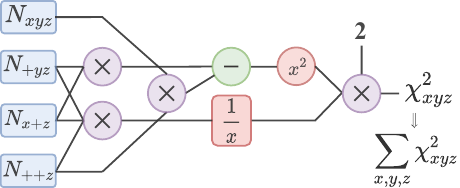}
    \caption{The $\chi^2$ Calculation Circuit.}
    \label{fig_chi2_circuit}
\end{figure}

\del{
\begin{figure*}[h]
    \centering
    \includegraphics[width=0.8\linewidth]{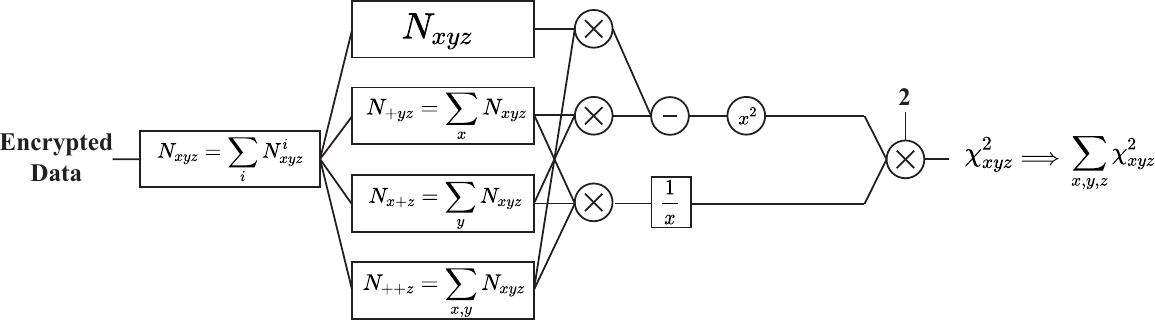}
    \caption{\ZYW{The execution circuit of }$\chi^2$}
    \label{fig_chi2_circuit}
\end{figure*}
}

\subsubsection{Calculations for $G^2$ Test}
Here, we discuss the computation process of $G^2$, which also faces the same problem that has two reciprocal operations.
We adopt a similar solution to simplify the equation by multiplying $N_{++z}$ to the denominator and numerator of $\frac{N_{xyz}}{E_{xyz}}$ together. In the end, we obtain a simpler equation that only requires one reciprocal operation. The derivation is shown as follows:

\begin{equation*}
\begin{aligned}
    G^2 &= 2 \sum_{x, y, z} N_{xyz} \log \frac{N_{xyz}}{E_{xyz}} = 2 \sum_{x, y, z} N_{xyz} \log \frac{N_{xyz}}{\frac{N_{x+z} N_{+yz}}{N_{++z}}} \\
    &= 2 \sum_{x, y, z} N_{xyz} \log \frac{N_{xyz}N_{++z}}{N_{x+z}N_{+yz}}
\end{aligned}
\end{equation*}

Figure~\ref{fig_g2_circuit} describes details of the arithmetic circuit. $N_{+yz}$, $N_{x+z}$ and $N_{++z}$ can be obtained from $N_{xyz}$. Then, they are fed into the circuit shown in the figure. After that, we calculate $N_{xyz}N_{++z}$ and $\frac{1}{N_{x+z}N_{+yz}}$ simultaneously and multiply them to get $\frac{N_{xyz}N_{++z}}{N_{x+z}N_{+yz}}$. $\frac{N_{xyz}N_{++z}}{N_{x+z}N_{+yz}}$ is input to the $\log$ module and multiply with $N_{xyz}$ to obtain $G^2$ finally.

\begin{figure}[hb]
    \centering
    \includegraphics[width=0.8\linewidth]{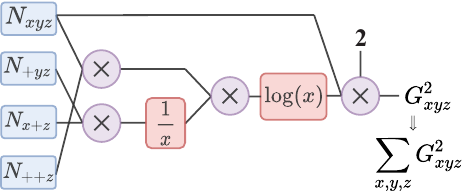}
    \caption{The $G^2$ Calculation Circuit.}
    \label{fig_g2_circuit}
\end{figure}

\subsubsection{Enhancing Efficiency with Batching}
It is known that fully homomorphic encryption is computationally expensive, and parallelism is a technique to accelerate computation. However, the crucial path in the circuit cannot be executed concurrently due to data dependency, while the crucial path dominates the total execution time. Therefore, it is infeasible to decompose a circuit and runs its subparts in parallel. In this paper, we utilize multiple circuit parallelism to improve efficiency. Following the CKKS scheme, a batch of input values can be encoded in one vector and be performed as one input to participate in operations.
After finishing these computations, the outputs are also formed as a vector that contains the resulting values. 
The batching scheme is defined as follows:
\begin{align*}
&\mathcal{I} = \{I_1 \oplus I_2\oplus...\oplus I_n\},\\
&\mathcal{O} = \{O_1\oplus O_2\oplus ...\oplus O_n\},\\
&f(\mathcal{I}) = \mathcal{O}, f(I_i) = O_i,
\end{align*}
where $f:\mathcal{I}\rightarrow \mathcal{O}$ is the arithmetic circuit; $I_i$ and $O_i$ is the $i$-th input and output value, respectively; $\mathcal{I}$ and $\mathcal{O}$ is the input and output vectors that concat all the input and output values, respectively.
cAs can be seen from the above scheme, executing the arithmetic circuit for every input value one by one can obtain the same result as running the circuit in one batch. The batching technique can significantly enhance efficiency, thanks to its single instruction multiple data (SIMD) nature.
In our causal structure setting, one CI test may take a large amount of running time, which is much larger than the overhead of batching. 

\subsubsection{Other Optimization Details}
Here, we provide the details of other optimization techniques, including initial guesses selection for the reciprocal circuit, point of Taylor expansion, and packing ciphertext for efficient communication.

\begin{itemize}
    \item \textbf{Initial Guess}: The initial guess of the reciprocal should be selected carefully for fast convergence. For the input $x$, the initial guess should be smaller than its reciprocal value~\cite{kornerup2003choosing}. Otherwise, it may not converge to the correct result. However, overly small initial guesses lead to excessive iterations and slow convergence. In this paper, we propose to choose the initial guess as the reciprocal of the maximum input, which guarantees correctness while achieving good efficiency. 
    \item \textbf{Expansion Point}: The point of Taylor expansion $x_0$ is crucial, because the approximate value is more precise when input value $x$ is near $x_0$. It means that $x_0$ should ideally be in the center of the domain of $x$.
    For $\log \frac{N_{xyz}}{E_{xyz}}$, it is common to compute it by ($\log N_{xyz} - \log E_{xyz}$). However, it is difficult to find where to expand $\log N_{xyz}$ and $\log E_{xyz}$, because values of $N_{xyz}$ and $E_{xyz}$ vary. In comparison, $\frac{N_{xyz}}{E_{xyz}}$ is around $1$, since $N_{xyz}$ and $E_{xyz}$ have similar values in most cases. Therefore, expanding the function in $x_0 = 1$ leads to a more accurate solution for calculation. 
    \item \textbf{Ciphertext Condensation}: For computation efficiency, ciphertext occupies a fixed space in memory at the beginning. The subsequent growth of ciphertext due to the progress of computation does not need to allocate new space, thanks to the memory pre-allocation. However, to perform efficient communication, we condense the unused part of the ciphertext and only transfer the used part to the receiver. Then, the receiver rebuilds the ciphertext by compositing the used part.
\end{itemize}
\subsection{Learning with Differential Privacy}
We describe the differential privacy module in this subsection.
Considering the $G^2$ computation in the distributed setting, it is easy to steal information in the data transfer process. 
Attackers analyze local results from each machine and infer individual information. To avoid the inference that obtains local information, we adopt differential to enhance the security of data transfer and aggregation.

We can recap Lines 10-12 of Algorithm~\ref{alg_disjoint_pc_stable}, which sends and receives the plaintext of local results, leaking local information about the individual privacy of machines.
To protect the privacy of transfer and processing, a simple approach is to add noise that the total sum is zero to these local results.  
One approach to achieve $\epsilon$-DP is to draw noise from a Laplace distribution $Lap(0, \frac{\Delta f}{\epsilon})$ that the expected mean is zero and scale is $\frac{\Delta f}{\epsilon}$~\cite{dwork2006differential}. Equation~\eqref{eq_Laplace} shows the Laplace Mechanism function $F(\mathcal{D})$, which can add noise to the original function $f(\mathcal{D})$. $\Delta f = \max_{\mathcal{D'}} || \mathcal{D} - \mathcal{D'} ||$, where $\mathcal{D'}$ only has one-sample difference with $\mathcal{D}$.
\begin{equation}
\label{eq_Laplace}
    F(\mathcal{D}) = f(\mathcal{D}) + Lap(0, \frac{\Delta f}{\epsilon}).
\end{equation}

In the context of causal structure learning, we modify the four local results $N_{xyz}^i$, $N_{x+z}^i$, $N_{+yz}^i$, and $N_{++z}^i$ to  $F(N_{xyz}^i)$, $F(N_{x+z}^i)$, $F(N_{+yz}^i)$, and $F(N_{++z}^i)$. The modified local results can obtain an approximate result of the original result. In the training process, the local results require noise to protect the information. Large noise is beneficial to protect privacy but is harmful to the original information. How to control the size of the noise is also a crucial problem, and one needs to balance privacy and accuracy problems.

\begin{table}
    \centering
    \caption{Dataset information.}
    \label{tab:dataset}
    \begin{tabular}{ccccc}
        \toprule
        Dataset & \#Nodes & \#Edges & \#States  & \#Samples\\ 
        \midrule
        \textit{child} & 20 & 25 & 6  &5000\\ 
        \textit{insurance} & 27 & 52 & 5  &5000\\ 
        \textit{water} & 32 & 66 & 4  &5000\\ 
        \textit{alarm} & 37 & 46 & 4  &5000\\ 
        \textit{hepar2} & 70 & 123 & 4  &5000\\ 
        \textit{win95pts} & 76 & 112 & 2 &5000\\ 
        \bottomrule
    \end{tabular}
    
\end{table}

\section{Experiment}
\label{sec_exp}
\del{In this section, we conduct experiments to study the efficacy and the performance of our differential privacy (DP) and fully homomorphic encryption (FHE) implementation and compare the results with baseline methods. In addition, we analyze the elapsed time, memory consumption, and communication cost of our proposed method. Finally, we discuss tradeoffs among execution time, memory consumption, and precision.}

In this section, we undertake experiments to investigate the efficacy and performance of our implementation that incorporates differential privacy (DP) and fully homomorphic encryption (FHE), and we compare the findings with baseline methods. Furthermore, we analyze the elapsed time, memory usage, and communication overhead associated with our proposed method. Finally, we investigate the trade-off between execution duration, memory utilization, and precision.

\subsection{Experiment Setting}
The experiments were conducted in 10 machines that are equipped with AMD EPYZ 7502 32 cores-64 threads CPU and 320GB memory. The proposed causal structure learning algorithm is implemented in C++ under the C++20 standard, with OpenMPI to support distributed and parallel computing. To implement fully homomorphic encryption computation, we adopt the Microsoft SEAL~\cite{sealcrypto} as FHE library and EVA~\cite{10.1145/3385412.3386023} as FHE compiler. We used $\chi^2$ and $G^2$ test statistics to perform the CI tests while setting the significance level $\alpha$ to 0.05 in all the experiments.

\textbf{Datasets}: The datasets used in the experiments are summarized in Table~\ref{tab:dataset}. These datasets used are generated from eight benchmark causal structures of different sizes and max status. These datasets illustrate problems from different fields and are widely used in other causal structure learning studies~\cite{scutari2014bayesian}. We obtained 5,000 samples of data with no missing values from each benchmark causal structure. Each dataset has nodes, edges, and states, which composite a causal structure. 
Every node represents one variable in a dataset, which is connected with other variables by edges. One edge between two nodes indicates one conditional dependency relation between two variables with an arrow for direction. States represent values of variables and are categorical types. Two-state variables have two types of value: $True$ and $False$. Three-state may have $High$, $Middle$, and $Low$. Edges from one node may arrive at different nodes in different states.
For example, there are three nodes $\mathcal{V}_1$, $\mathcal{V}_2$, and $\mathcal{V}_3$ under a two-state setting. Edge $\mathcal{V}_1 \xrightarrow{True} \mathcal{V}_2$ means that when value of $\mathcal{V}_1$ is $True$, next node of $\mathcal{V}_1$ is $\mathcal{V}_2$. In addition, there may also have one edge $\mathcal{V}_1 \xrightarrow{False} \mathcal{V}_3$ or $\mathcal{V}_1\xrightarrow{False} \mathcal{V}_2$. If $\mathcal{V}_1 \xrightarrow{True} \mathcal{V}_2$ and $\mathcal{V}_1\xrightarrow{False} \mathcal{V}_2$, then it represents that both cases of $\mathcal{V}_1$ taking $True$ or $False$ transit to $\mathcal{V}_2$.

\textbf{Distributed Learning Setting}: For the $k$ machines setting, we horizontally partition dataset $\mathcal{D}$ into $k$ parts; $\mathcal{D}_i$ represents the $i$-th dataset and is the dataset of $i$-th machine $\mathcal{M}_i$. 
We set $k=2$ for most cases except the scalability part of experiments. 
For each level of causal structure learning, we encode all CI tests in one batch and input the batch to the arithmetic circuit in one pass to obtain encrypted p-values.
Then the encrypted p-values are sent to machines and machines decrypted p-values to decide the existence of edges.
Finally, each machine obtains consistent causal structures.
For each level in causal structure learning, we packed all the CI tests in one batch on the client node and passed the encrypted batch to the arithmetic circuit in the coordinator in one pass to obtain encrypted $\chi^2$ or $G^2$ values. Once the values are sent back to the client machines, they are decrypted and the p-value is obtained from the lookup table. Finally, the dependence of each edge can be determined, by which consistent causal structures on the client machine are obtained.

\begin{table*}
    \centering
    \caption{Consistency of CI test at each level between our approach and the plain version.}
    \label{tab_consist}
    \begin{tabular}{cllllllllllll}
    \toprule
        Dataset & \multicolumn{2}{c}{\textit{win95pts}}  & \multicolumn{2}{c}{\textit{alarm}}  & \multicolumn{2}{c}{\textit{hepar2}}  & \multicolumn{2}{c}{\textit{water}}  & \multicolumn{2}{c}{\textit{insurance}} &\multicolumn{2}{c}{\textit{child}} \\ 
        \midrule
        Level & $\chi^2$ & $G^2$ & $\chi^2$ & $G^2$ & $\chi^2$ & $G^2$ & $\chi^2$ & $G^2$ & $\chi^2$ & $G^2$ & $\chi^2$ & $G^2$ \\ 
        \midrule
        0 & 0.931  & 0.989  & 0.881  & 0.991  & 0.864  & 0.993  & 0.925  & 0.988  & 0.937  & 0.972  & 0.911  & 1.000  \\ 
        1 & 0.731  & 0.846  & 0.833  & 0.896  & 0.611  & 0.833  & 0.849  & 0.792  & 0.865  & 0.814  & 0.928  & 0.963  \\ 
        Average & 0.746  & 0.833  & 0.833  & 0.900  & 0.641  & 0.863  & 0.854  & 0.833  & 0.862  & 0.826  & 0.922  & 0.960 \\ 
        \bottomrule
    \end{tabular}
\end{table*}

\subsection{Consistency Verification}
The most important object for privacy-preserving causal structure learning is correctness, which means that we should ensure the proposed method produces similar results as the one without encryption. The baseline of this experiment is the causal structure learning in plaintext method discussed in Section~\ref{paper:plaintext}. We used the consistency of CI tests, and compared two causal structures with structural hamming distance (SHD). The SHD measurements are commonly used in causal structure learning~\cite{bookstein2002generalized}. The experiments were conducted under $G^2$ and $\chi^2$ tests for the learned causal structure to verify the correctness.

\begin{figure}[h]
	\centering  
        \subfloat[CI test]
	{\includegraphics[width=0.49\linewidth]{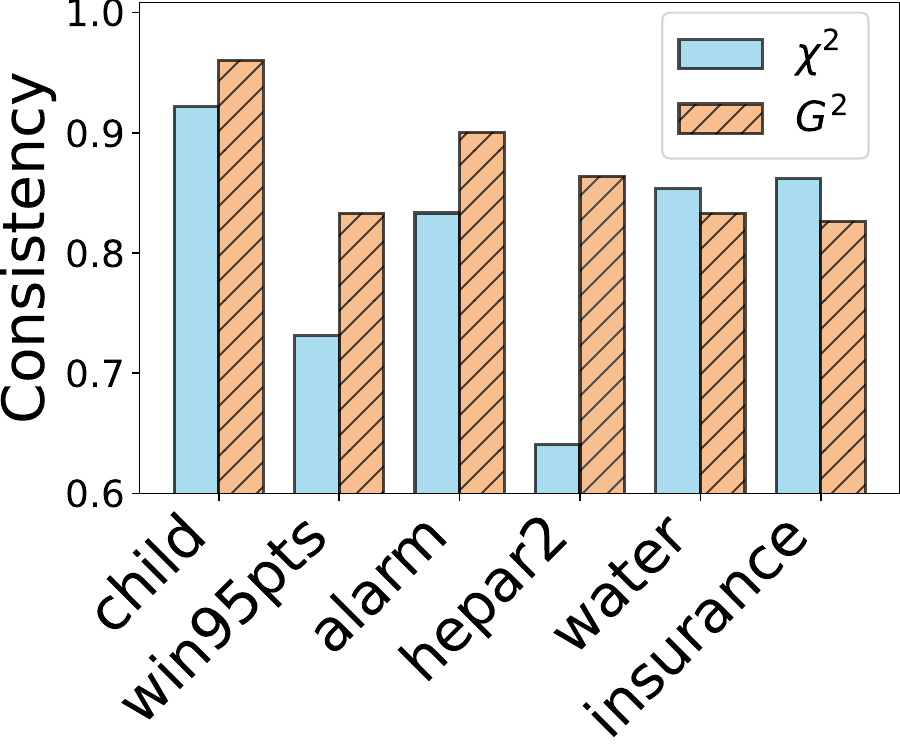}
		\label{fig_ci}} 
	\subfloat[SHD]
	{\includegraphics[width=0.49\linewidth]{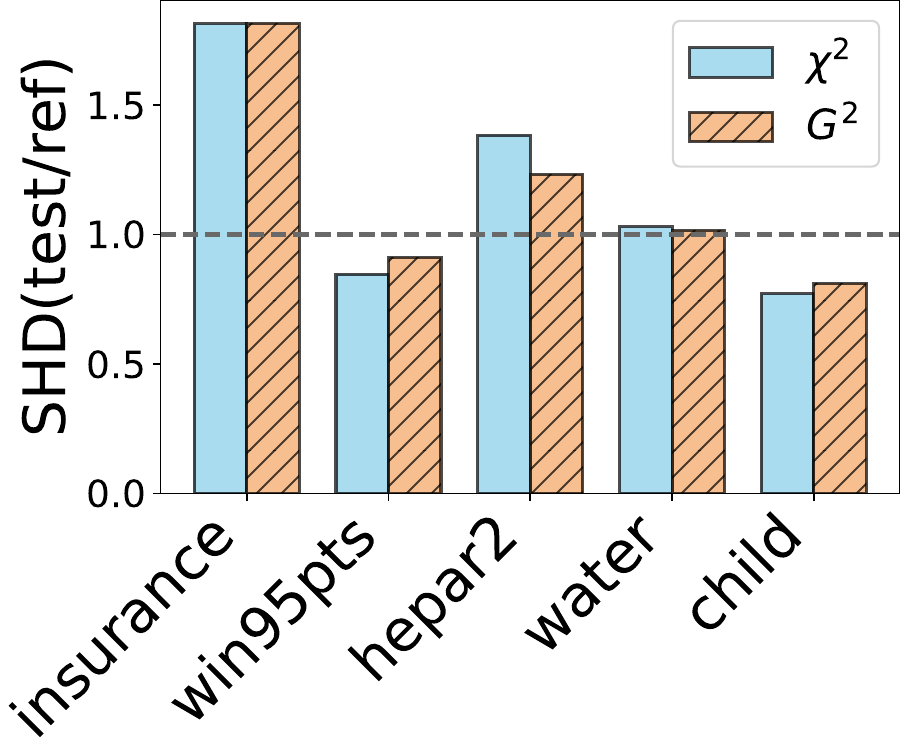}
		\label{fig_shd}} 
	\caption{SHD and CI tests consistency.}
	\label{fig_shd_and_ci} 
\end{figure} 
For the CI tests, we compared the results output by our method and the plaintext method, and then counted the consistent numbers.
Figure~\ref{fig_ci} and Table~\ref{tab_consist} illustrate the CI test consistency ratios: the overlapping ratio of CI test outcomes from our method and the plaintext method. From Table~\ref{tab_consist}, the average results across 6 datasets demonstrate that the consistency of our method is around 85\%. An observation is that in $G^2$ of \textit{alarm} and $\chi^2$ of \textit{child}, the consistency is over 90\%. In \textit{win95pts}, \textit{alarm}, \textit{hepar2}, \textit{insurance}, and \textit{child}, the $G^2$ obtains better results than $\chi^2$, while results of $\chi^2$ are better in \textit{water}.

From the level perspective, the CI tests at level 0 are more consistent than those at level 1, because level 0 does not have a separating set, which leads to shorter computation process, which has less precision loss than a longer process. From different tests perspective, the CI tests based on $\chi^2$ is more consistent than those based on $G^2$ in level 0, but worsen at level 1. This is because square operations have better precision and more computation requirement. At level 0, it has high precision due to sufficient resources, while it is not affordable for the square explosion and leads to precision loss at level 1. 
In contrast, $G^2$ adopts $\log$ to reduce the amount of test value. Although $G^2$ is not accurate enough at level 0 (about 10\% less than $\chi^2$), it slightly loses precision at level 1. This is because the number of CI tests at level 1 is much larger than that at level 0, according to the combination number, leading to better overall performance in $G^2$.

We also verify the structure information through the ratio of SHD for the reference structures (produced by plaintext version) and our method's. Figure~\ref{fig_shd} shows that our method can obtain a similar structure to the reference one. In \textit{win95pts}, \textit{water}, \textit{hepar2}, and \textit{child}, the ratio is mildly small than 1, indicating highly similar results. In \textit{insurance}, our results are better than references, because the batching technique checks all the CI tests simultaneously while the plaintext version conducts CI tests one by one. 
Considering all the CI tests in one level together benefits the performance of final results due to its comprehensive decision-making for each CI test and edge. 
Our method looks through these CI tests and finds the highest p-value CI test for each edge, which produces the best separating set for the edge orientation step.

\subsection{Execution Time Analysis}
\label{paper:ex_time}
The execution time of our method is also a crucial concern in the experiments.
We measured total elapsed times over 6 datasets and recorded them in Figure~\ref{fig_total_time}. 
It is easy to find that the $\chi^2$ test executes faster than the $G^2$ test since the arithmetic circuit of $\chi^2$ is simpler than the arithmetic circuit of $G^2$.
More complex circuits require longer execution times for computation.

Observing elapsed times for different datasets, \textit{win95pts} is the fastest, and \textit{child} is the slowest. 
At first glance, this result seems counter-intuitive, as \textit{win95pts} has the most nodes and the second most edges, while \textit{child} has the fewest nodes and edges. The key reason is that elapsed time is only affected by the number of states and increases exponentially with the increment of the number of states. Figure~\ref{fig_dim_time} studies elapsed time of one batch of $2^{15}$ CI tests in different numbers of states. 
The workload of this batch is sufficient to evaluate the effects of the number of states.
As we can see from the figure, elapsed time grows up with the incremental of states in both $G^2$ and $\chi^2$. $G^2$ costs more time than $\chi^2$ at each number of states. When the number of states is 6, the elapsed time of $G^2$ and $\chi^2$ is 4 times larger than their elapsed time when the number of states is 2.

\begin{figure}[h]
	\centering
	\subfloat[Total elapsed time]
	{\includegraphics[width=0.45\linewidth]{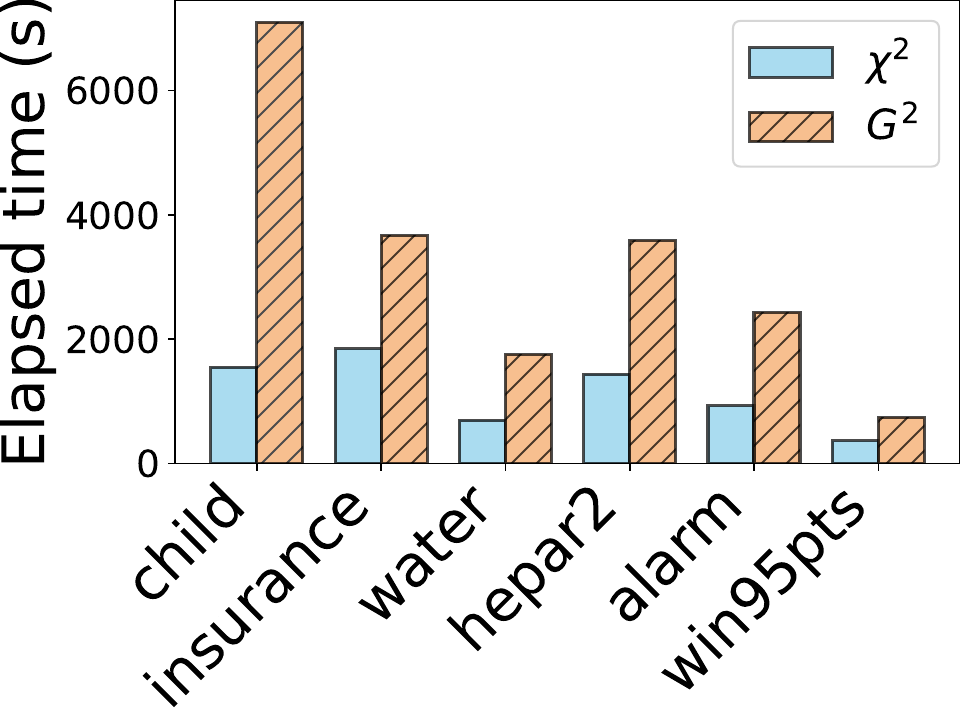}
		\label{fig_total_time}} 
	\subfloat[Elapsed time for each part]
	{\includegraphics[width=0.45\linewidth]{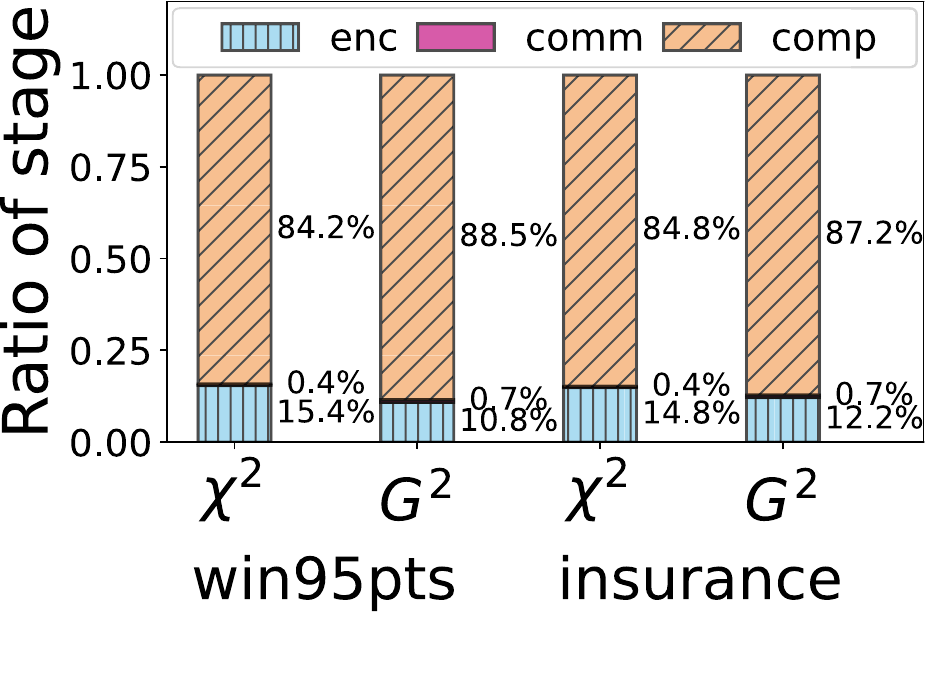}
		\label{fig_time_ratio}} 
	 \caption{Time comparison. (a) is the total time for $\chi^2$ and $G^2$. (b) indicates each part contributing to elapsed time, where ``comm", ``enc", and ``comp" represent communication, encryption, and computation respectively.}  
	\label{fig_time} 
\end{figure} 

We also study the time consumption in different parts in Figrue~\ref{fig_time_ratio}. 
In each dataset, our method spends above 84\% time on computations, around 10\%-15\% time on encryption, and less than 1\% time on communication. 
Most of the elapsed time is on computations, which feed encrypted inputs into arithmetic circuits and obtain encrypted outputs.
A small part of the time is used for encryption, which encrypts inputs and decrypts encrypted outputs.
The time for communication is omittable, because of its faint effect on the total time.
Here $G^2$ still puts more time into computation than $\chi^2$ as the total elapsed time shown in Figure~\ref{fig_total_time}.

In Table~\ref{tab:compare_plaintext}, we compare the elapsed time of our method with the plaintext version. Although the plaintext version finishes the execution in 5 seconds, our method is practical to complete running in 6 to 30 minutes under privacy protection. Note that causal structure learning is an offline process, and hence 30 minutes of learning time is often acceptable.

\begin{table}[h]
    \centering
    \caption{Elapsed time (sec) comparison in $\chi^2$ between our method and the plaintext version.}
    \begin{tabular}{ccc}
        \toprule
        Dataset & Plaintext & Ours  \\ 
        \midrule
        \textit{child} & 0.18 & 1545 \\ 
        \textit{insurance} & 0.24 & 1845 \\ 
        \textit{water} &  0.07 & 698 \\ 
        \textit{alarm} & 0.12 & 933\\ 
        \textit{hepar2} & 1.57 & 1433\\ 
        \textit{win95pts} & 0.51 & 366 \\ 
        \bottomrule
    \end{tabular}
    \label{tab:compare_plaintext}
\end{table}

\begin{figure}
    \centering
	\subfloat[Elapsed time]
	{\includegraphics[width=0.49\linewidth]{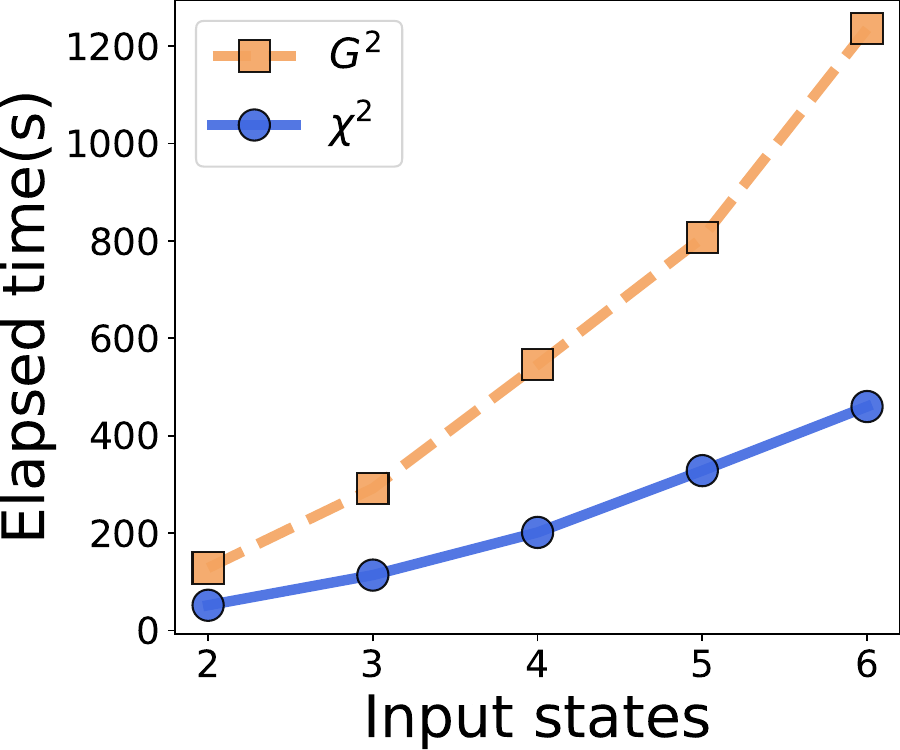}
		\label{fig_dim_time}} 
	\subfloat[Memory usage]
	{\includegraphics[width=0.49\linewidth]{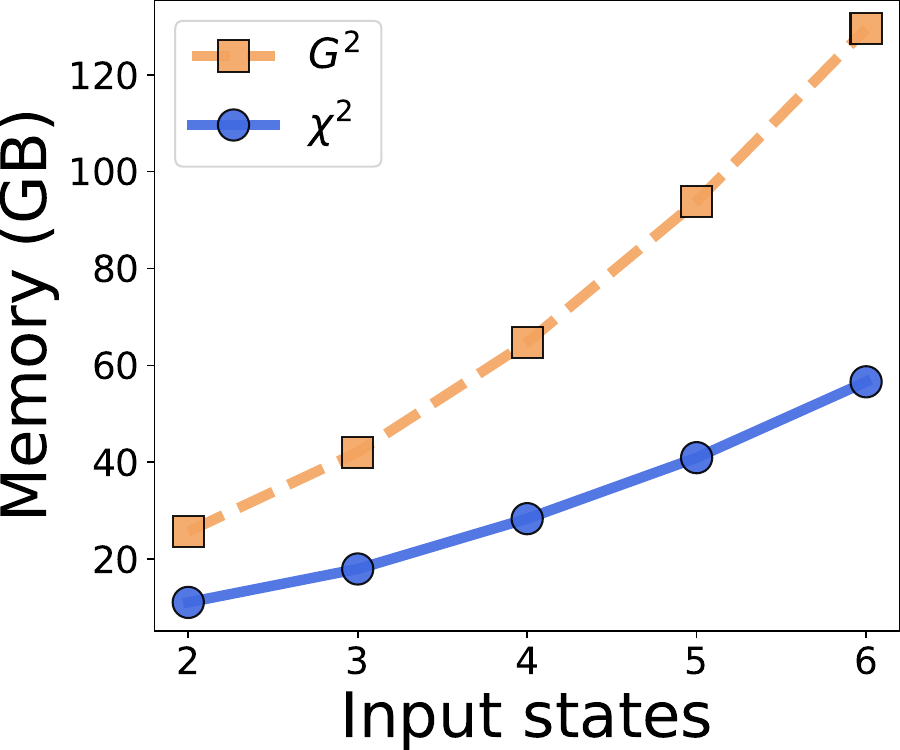}
		\label{fig_dim_mem}} 
	 \caption{Time and memory for a large batch with $G^2$ and $\chi^2$.}  
	\label{fig_dim} 
\end{figure}

\subsection{Memory and Communication Analysis}
In this subsection, we study memory usage and communication overhead in our method. We test the maximum memory usage for executions on different datasets in $G^2$ and $\chi^2$, and then record results on Figure~\ref{fig_overall_memory}. 
It is to conclude that $G^2$ consumes more memory than $\chi^2$, due to its more complex arithmetic circuit.
To build a more complex circuit, the program needs to allocate more memory to buffer it.
In addition, the number of nodes or edges does not affect the memory consumption of the proposed method, since \textit{win95pts} has the most edges and nodes but costs the least memory. 
In fact, the only factor for memory usage is the number of states, similar to the phenomenon in elapsed time studied in Section~\ref{paper:ex_time}. 
We find \textit{win95pts} has 2 states and uses the least memory; \textit{alarm}, \textit{hepar2}, and \textit{water} have 4 states that get the same memory usage. The \textit{child} dataset owns 6 states, and consumes the most memory.
To further verify our conclusion, we test the memory usage for a batch of $2^{15}$ CI tests with $G^2$ and $\chi^2$ in Figure~\ref{fig_dim_mem}. Following the increase of states, memory usage shows similar results as the elapsed time shown in Figure~\ref{fig_dim_time} on the different datasets.

\begin{figure}
    \centering
	\subfloat[Maximum memory usage]
	{\includegraphics[width=0.49\linewidth]{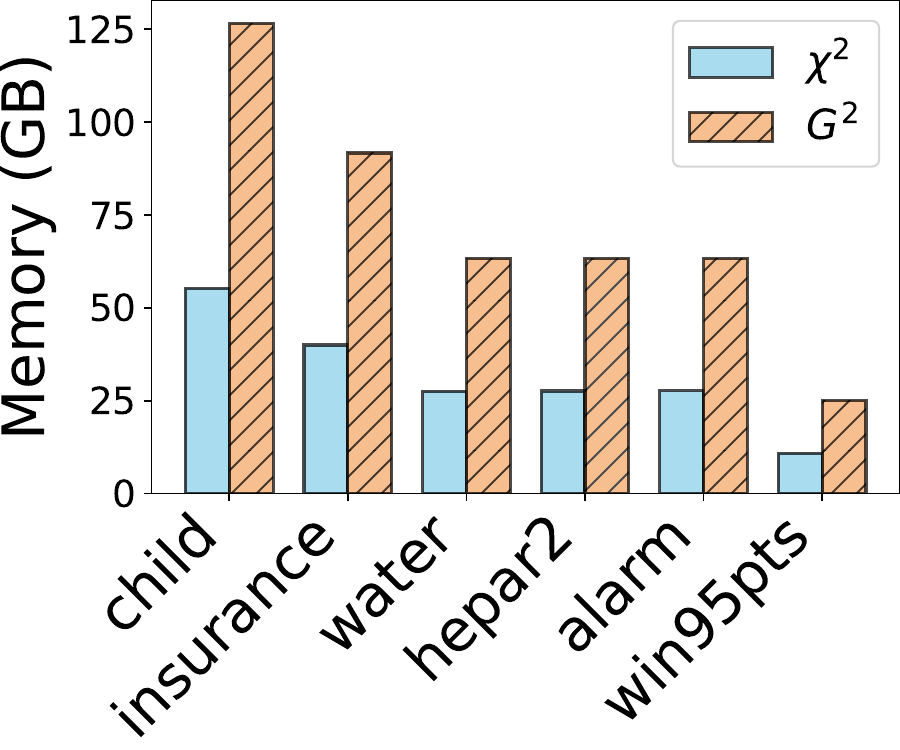}
		\label{fig_overall_memory}} 
	\subfloat[Communication cost]
	{\includegraphics[width=0.49\linewidth]{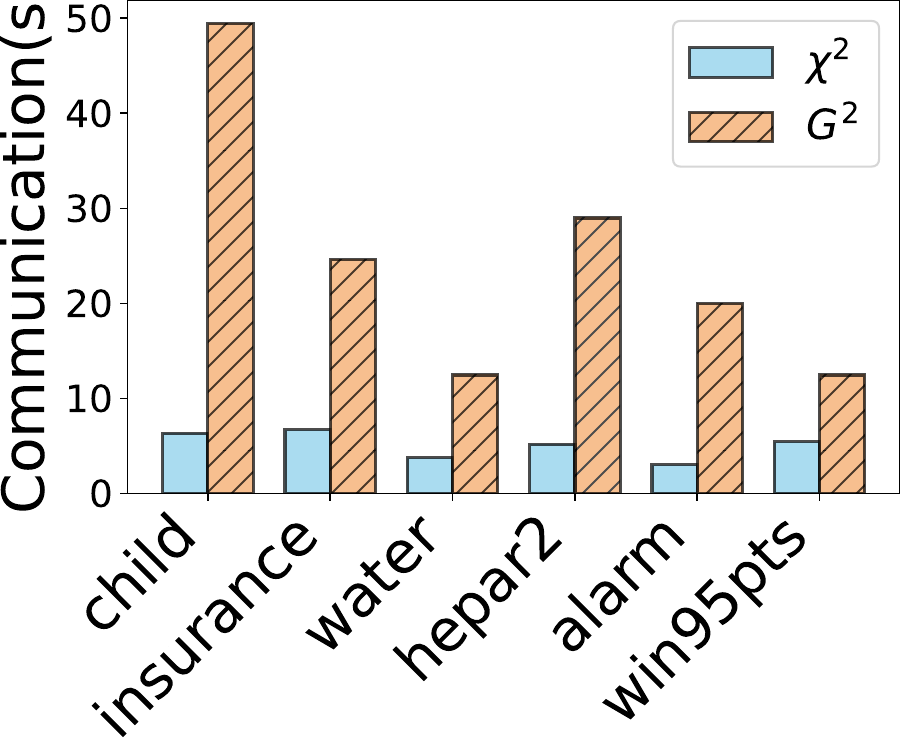}
		\label{fig_overall_comm}} 

  \caption{Max memory usage and communication time for different datasets in $\chi^2$
and $G^2$.}  
	\label{fig_comm_and_memory} 
\end{figure}

We measure the communication time for different datasets in $G^2$ and $\chi^2$ and record results in Figure~\ref{fig_overall_comm}. 
The communication time is still related to the number of states. 
\textit{win95pts} use the least time on communication while \textit{child} uses the most time on communication. For datasets that have the same number of states, such as \textit{water}, \textit{alarm} and \textit{hepar2}, more edges and nodes lead to longer communication time, due to more CI tests for more edges and larger separating sets for more nodes. $G^2$ also spend more time on communication than $\chi^2$ because the size of ciphertext of $G^2$ is much bigger than that of $\chi^2$. 

We also estimate the memory usage at different time slots in \textit{insurance} with $\chi^2$ as an example, detailed in Figure~\ref{fig_memory_level}, revealing three stages. The first part conducting data-loading sees an initial memory spike to 20GB. The second, data encryption and communication, entails a sharp memory increase to 45GB due to encrypted data's high memory demands. The final computation stage displays steady memory usage at 45GB because of the computation graph's occupation.

\begin{figure}[h]
    \centering
    \includegraphics[width=0.9\linewidth]{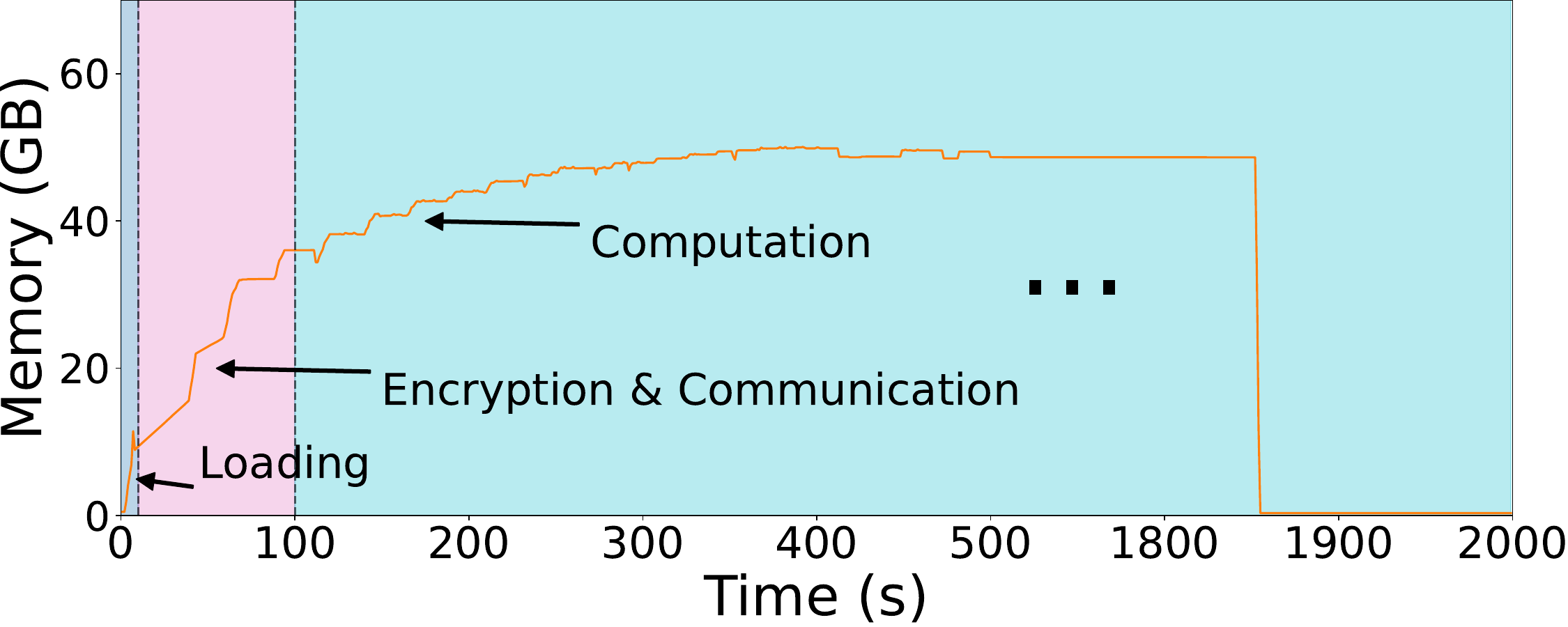}
        \caption{Memory consumption during learning.}
    \label{fig_memory_level}
\end{figure}

\subsection{Hyperparameter Study}
Hyperparameters are crucial configurations in our method. We study three important hyperparameters in this subsection.
First, we discuss the effect of batching size on elapsed time and memory usage. 
Then, we analyze the times of iterations for the Newton-Raphson reciprocal circuit and the series for Taylor expansion.
\begin{figure}
    \centering
    \subfloat[Elapsed time]
	{\includegraphics[width=0.46\linewidth]{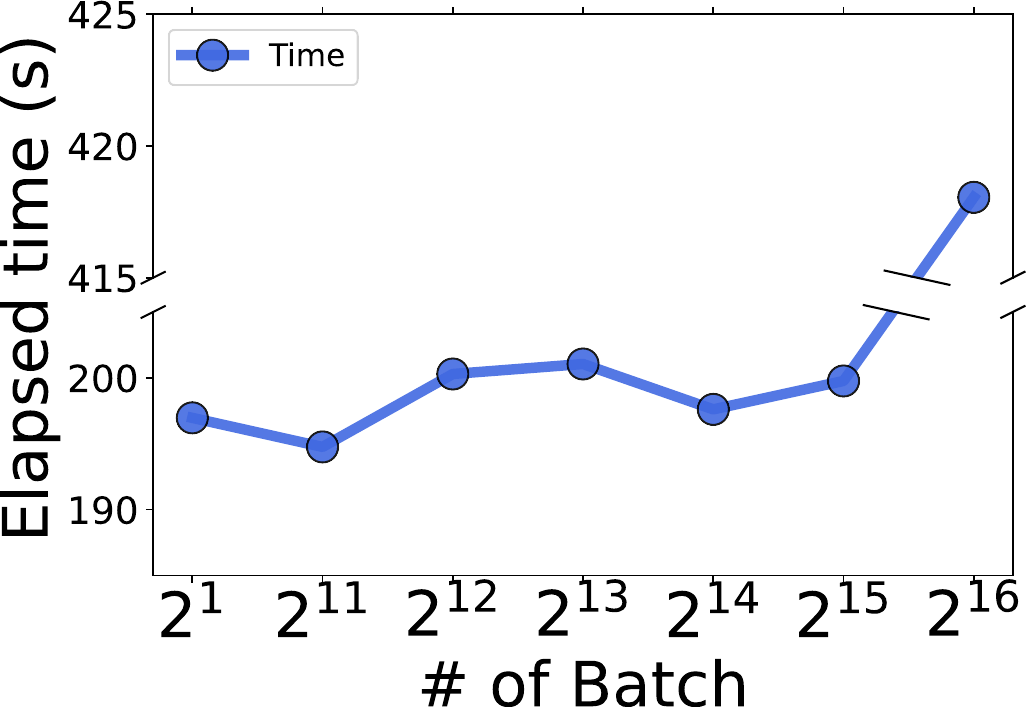}
		\label{fig_batching_shd}} 
	\subfloat[Memory usage]
	{\includegraphics[width=0.46\linewidth]{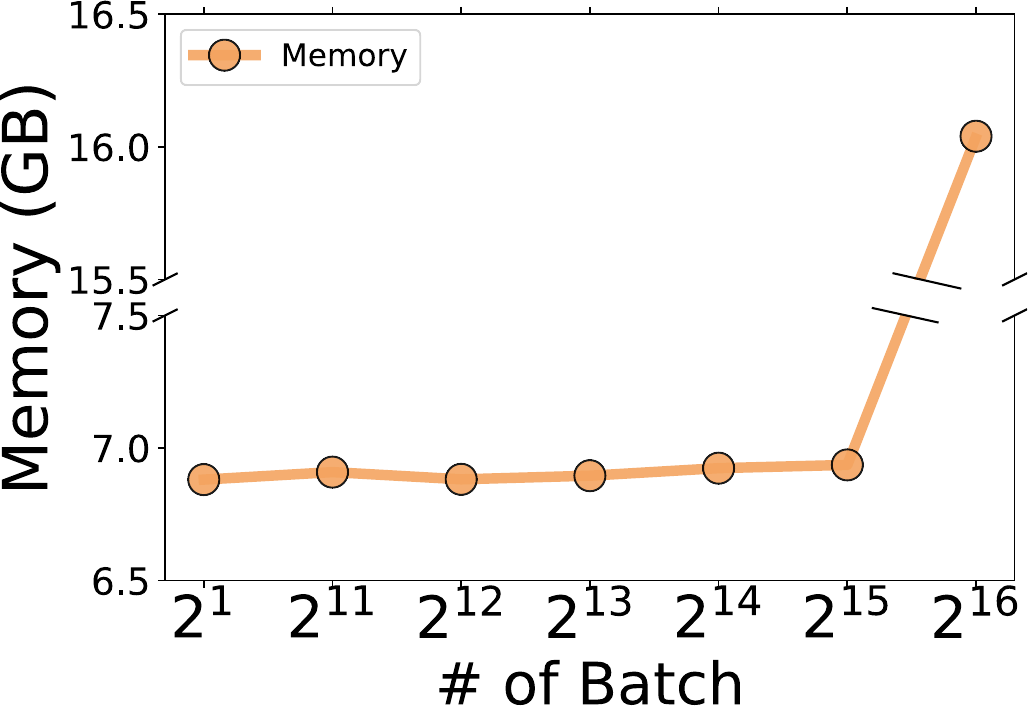}}
    \caption{Time and memory for $2^1$ to $2^{16}$ elements in batch.}
    \label{fig_batching}
\end{figure}

We test the memory usage and elapsed time in elements in one batch from $2^1$ to $2^{16}$ and show the result in Figure~\ref{fig_batching}.
When the number of elements in one batch is not larger than $2^{15}$, the memory usage and elapsed time are kept the same as those of $2^1$, indicating that computing $2^1$ elements has the same overhead as computing $2^{15}$. Therefore, this parallel technique has $2^{14}$ times speedup.

For the Taylor expansion circuit, the relationship between elapsed time and memory usage in different numbers of series is shown in Figure~\ref{fig_log_exp}. We observe that the 3-series expansion has a better balance between precision and speed. 
For the Newton-Raphson reciprocal circuit, we varied the number of iterations and recorded the results shown in Figure~\ref{fig_rcp_exp}. As we can see from the result, having 10 iterations is a good tradeoff between precision and efficiency. 
\begin{figure}[h]
	\centering
	\subfloat[$\log(x)$ Taylor expansion]
	{\includegraphics[width=0.5\linewidth]{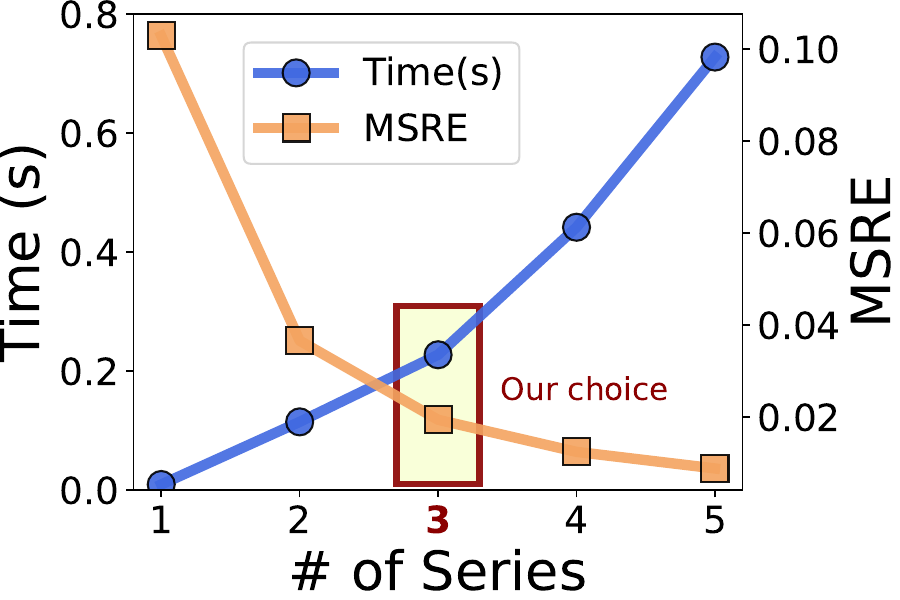}
		\label{fig_log_exp}} 
	\subfloat[$1/x$ Newton-Raphson reciprocal]
        {\includegraphics[width=0.5\linewidth]{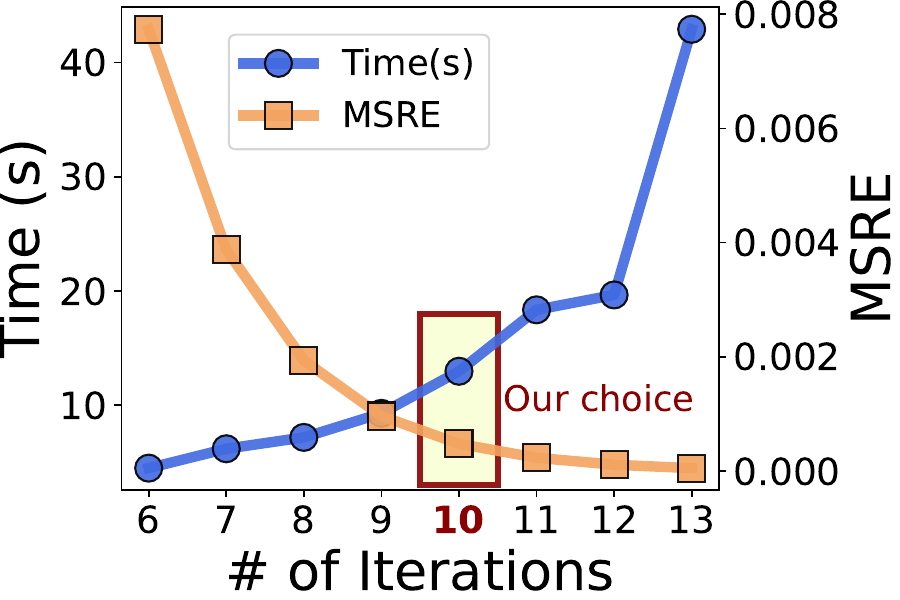} 
		\label{fig_rcp_exp}}
	\label{fig_arithmtic_appro} 
  \caption{Arithmetic circuit hyperparameter testing. Comparing times of iterations for $\log (x)$ and series of Taylor expansion for $1/x$.}
\end{figure}

\subsection{Scalability}
We evaluate the scalability of our method by enumerating the number of machines from 2 to 10. Figure~\ref{fig_scale_elap_time} illustrates the total elapsed time and computation time for different numbers of machines, showing a small rise in total time and stability in computation time. 
The increasing total time is due to more communication time for more ciphertexts needing to be transmitted, which is confirmed by the growing communication size in Figure~\ref{fig_scale_comm}. 
The coordinator undertakes all computation workload that is not changed in different numbers of machines, so the computation time keeps stable. Figure~\ref{fig_scale_comm} also demonstrates that memory consumption rises linearly to the number of machines.

\begin{figure}
	\centering 
	\subfloat[Elapsed time]
	{\includegraphics[width=0.45\linewidth]{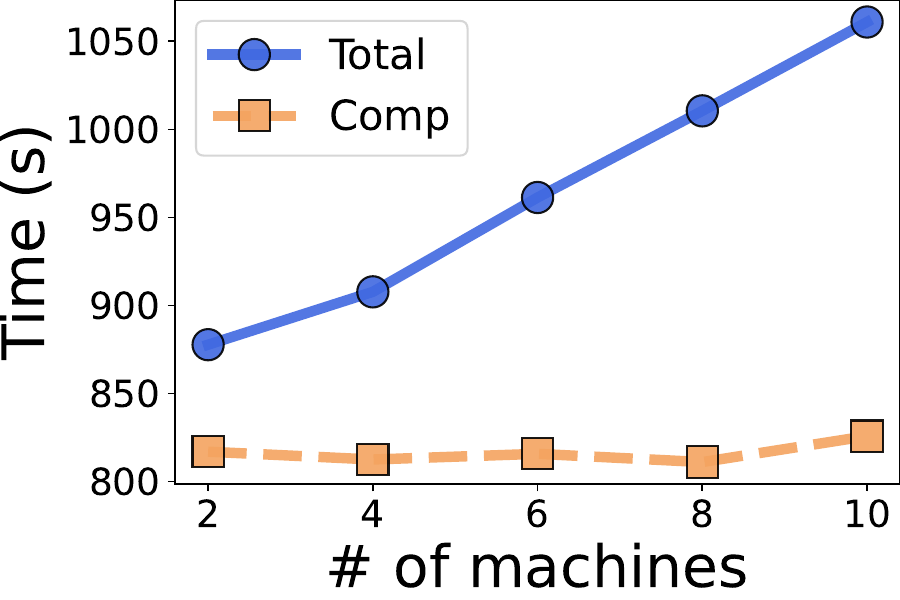}
		\label{fig_scale_elap_time}} 
	\subfloat[Communication and memory]
	{\includegraphics[width=0.45\linewidth]{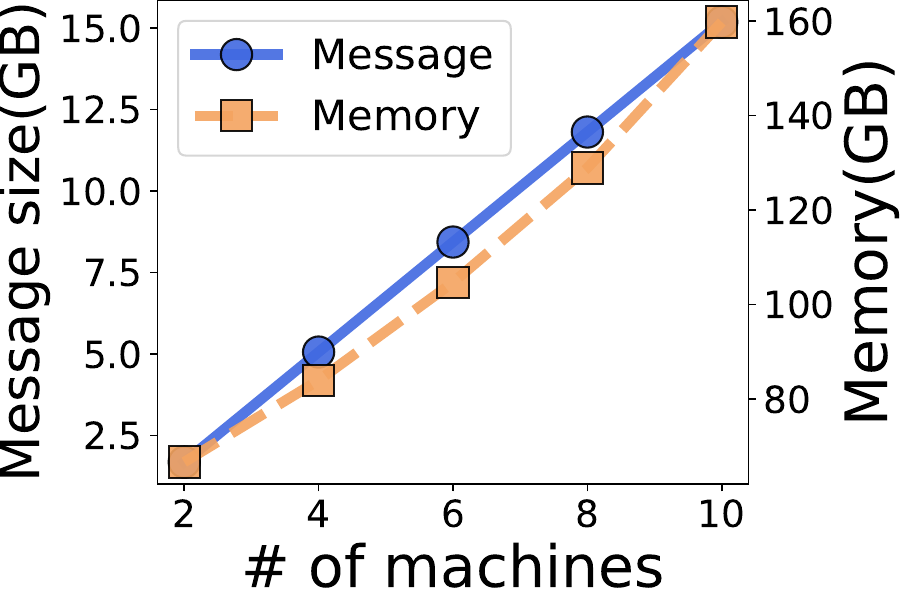} 
		\label{fig_scale_comm}} 
   \caption{Elapsed time, communication, and memory analysis for $2$ to $10$ machines.}
	\label{fig_scale} 
\end{figure}

\begin{figure}[ht]
    \centering
    \includegraphics[width=0.5\linewidth]{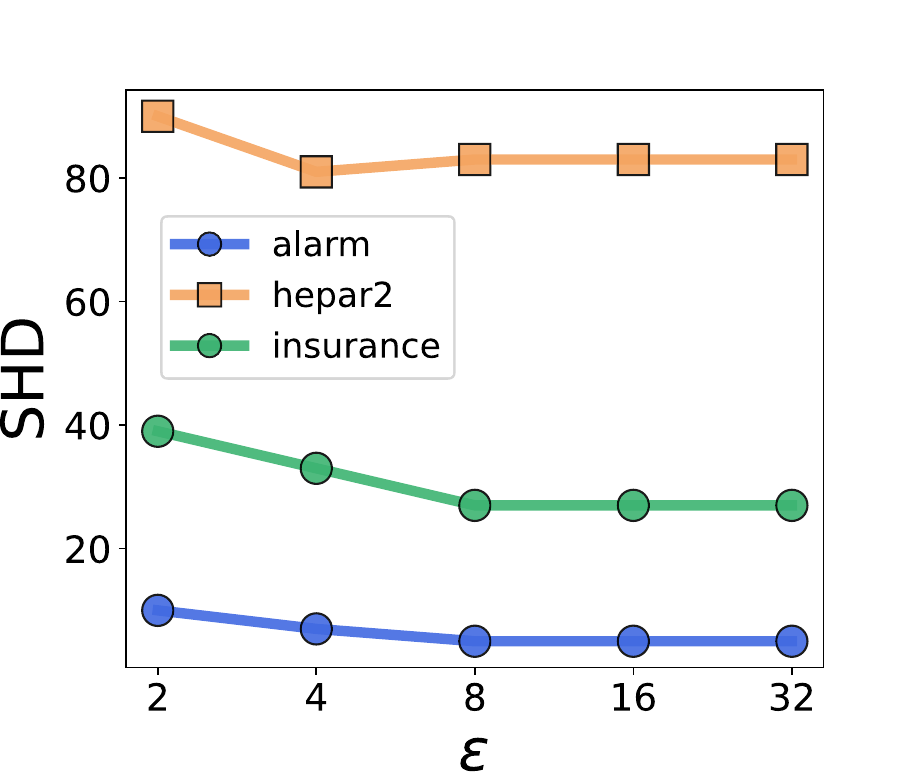}
    \caption{The SHD in three datasets in different $\epsilon$.}
    \label{fig_dp_overall}
\end{figure}

\subsection{Differential Privacy Experiments}
Here, we analyze our proposed method with differential privacy. The DP method only requires $O(1)$ time for privacy protection and has the same time complexity as the plaintext version, so we omit experiments of the elapsed time.
Figure~\ref{fig_dp_overall} shows the SHD results in three datasets: \textit{alarm}, \textit{herpar2}, and \textit{insurance} in different noise parameter $\epsilon$. 
The dash represents the SHD of the reference network produced by the plaintext method.  
Smaller $\epsilon$ leads to larger noise following the distribution $lap(0, \frac{\Delta f}{\epsilon})$. Therefore, the SHD of our method converges to the SHD of the plaintext method with the growth of $1/\epsilon$. 

\subsection{Experimental Summary}
Here, we summarize the key findings of our experiment. First of all, Our method achieves high CI test consistency and comparable structure with the plaintext version. Our method can be extended to differential privacy settings, which also achieves good quality precision compared with the plaintext version. Second, 84\% of the total time of our method is spent on computation, 10\%-15\% on encryption, and less than 1\% time on communication. Although the plaintext version finishes running in just 5 seconds, our method is still practical to complete learning causal structures in 6 to 30 minutes under privacy protection. Third, for different time slots in $insurance$, computation costs 45GB of memory, while encryption and communication consume 20GB to 45GB of memory. Communication overhead is relatively light in our proposed method, which is less than 50 seconds among datasets. Fourth, $2^{15}$ is the best number for batching, 3-term Taylor expansion, and 10-iteration reciprocal balance precision and efficiency. Finally, the total elapsed time, communication size, and memory usage increase linearly with the growth of the number of machines. The computation time remains constant when increasing the number of machines. 

\section{Related Work}
\label{paper:rw}
Here, we present the methodologies for causal structure learning, focusing on the enhancements achievable through distributed and parallel computing. Then, we explore the domain of privacy-preserving distributed learning algorithms in the context of causal structure learning.

\subsection{Causal Structure Learning}
\label{sec_related}

\del{Causal structures are powerful models for representation learning and reasoning under uncertainty in artificial intelligence. Causal Structures have recently attracted much attention within the research and industry communities. A crucial aspect of using causal structures is to learn the dependency graph of a causal structure from data, which is called \emph{structure learning}. In this paper, we categorize the related work on causal structure learning into two groups: score-based approaches and constraint-based approaches. }

Causal structures have emerged as potent models for representation learning and uncertainty management in the field of artificial intelligence, garnering significant interest within both research and industrial spheres. A key aspect of deploying causal structures involves determining the dependency graph from the data, a process referred to as \emph{structure learning}. This paper classifies the existing literature on causal structure learning into two primary categories: score-based and constraint-based methodologies.

\del{Constraint-based approaches~\cite{spirtes2000causation, colombo2014order, colombo2012learning, jiang2022fast} perform structure learning using a series of statistical tests, such as $\chi^2$ test, $G^2$ test, and mutual information test, to learn the conditional independence relationships among the variables. The DAG is then built according to these relations as constraints. Most of the constraint-based algorithms proceed along similar lines as the work of the PC-stable 
algorithm~\cite{spirtes2000causation, colombo2014order}. 
Unlike score-based approaches, it is generally non-trivial to perform algorithmic improvements for constraint-based approaches using general-purpose optimization theory. This paper mainly focuses on privacy-awareness of the constraint-based approaches in a distributed setting.}

\textbf{Constraint-based approaches}~\cite{spirtes2000causation, colombo2014order, colombo2012learning, jiang2022fast} undertake structure learning via a succession of statistical tests, such as the $\chi^2$ test, $G^2$ test, and mutual information test, to discern conditional independence relationships among variables. These relationships serve as constraints in building the DAG. The majority of constraint-based algorithms operate similarly to the PC-stable algorithm~\cite{spirtes2000causation, colombo2014order}. Unlike score-based approaches, enhancing constraint-based approaches using general-purpose optimization theory is inherently non-trivial. This paper primarily addresses the privacy-awareness of constraint-based approaches in distributed settings. 

\textbf{Score-based approaches}~\cite{acid2003searching, tian2013branch, myers2013learning} pursue the optimal DAG in accordance with scoring functions that assess the fitness of causal structures to the observed data. Commonly employed scores include BDeu, BIC, and MDL. Zhu et al~\cite{zhu2021efficient} have suggested a continuous optimization solution for recommendation systems. 

\del{Score-based approaches~\cite{acid2003searching, tian2013branch, myers2013learning} seek the best directed acyclic graph (DAG) according to scoring functions that measure the fitness of causal structures to the observed data. Widely adopted scores include BDeu, BIC, and MDL
. Zhu et al~\cite{zhu2021efficient} proposed a continuous optimization solution in recommendation systems. }

However, the number of possible DAGs is super-exponential to the number of variables~\cite{robinson1977counting}.
Hence, many score-based approaches employ heuristics, like greedy search or simulated annealing, in an attempt to reduce the search space. Such approaches can easily get trapped in local optima~\cite{scutari2019learns}. The optimization techniques in this paper focus on the constraint-based approaches which tend to scale better to high-dimensional data. 

\subsection{Privacy-Aware Distributed Learning}
\del{Due to frequent communication among machines, protecting sensitive information is a huge challenge in distributed learning tasks.
Sensitive information is easy to be leaked during inevitable data transfer and sharing progress, while attackers capture data packages to infer local or global information for malicious utilization. 
Some papers enhance the power of privacy protection in communication by adding noise in raw data~\cite{dwork2006differential}, sharing agreed functions for mutli-party~\cite{goldreich1998secure}, anonymous techinques~\cite{bonawitz2017practical}, or computation on ciphertexts without decrypting them~\cite{fontaine2007survey}. }

Preserving sensitive information in distributed learning tasks presents a significant challenge due to the necessity of frequent inter-machine communication. The inevitable data transfer and sharing process risk exposing sensitive information, which could potentially be intercepted by malicious parties seeking to infer local or global information. Several studies propose mitigating this risk by enhancing privacy protection in communication via the addition of noise to raw data~\cite{dwork2006differential}, sharing agreed functions for mutli-party~\cite{goldreich1998secure}, employing anonymization techniques~\cite{bonawitz2017practical}, or conducting computations on ciphertexts without decryption~\cite{fontaine2007survey}.

\del{There are two main privacy-aware techniques for distributed machine learning, including differential privacy (DP) and fully homomorphic encryption (FHE). The differential privacy (DP) distribution for privacy protection was used for composing local results to global~\cite{pathak2010multiparty}, which can avoid reverse reasoning personal data from global data movement. The DP algorithm can execute efficiently because it only requires adding noise to the original data. However, DP exposes plaintext to the main server, resulting in easily recovering the origin text~\cite{aono2017privacy}. }

\del{Fully Homomorphic Encryption (FHE)~\cite{fontaine2007survey} enables certain operations (such as addition or multiplication) to be performed directly on ciphertexts without requiring decryption, allowing computations to be performed on data while keeping its actual value unknown.
BatchCrypt~\cite{zhang2020batchcrypt} provides a system solution for distributed machine learning based on FHE that accelerates communication and encryption by batching data.}

Differential privacy (DP) and fully homomorphic encryption (FHE) are the primary privacy-aware techniques used in data management and distributed machine learning. DP has been used to compose local results globally~\cite{pathak2010multiparty}, thereby preventing the reverse-engineering of personal data from global data movements. Despite its efficiency, DP exposes plaintext to the main server, rendering the original text vulnerable to recovery~\cite{aono2017privacy}. Some works use DP to protect privacy in queries of database systems~\cite{cai2023privlava, dong2023better}. Conversely, FHE~\cite{fontaine2007survey} allows certain operations to be performed on ciphertexts without decryption, thereby preserving the confidentiality of the data. HEDA~\cite{ren2022heda} develops an FHE-based database analytical system combining to support SQL aggregation queries in encrypted databases. Radix-based parallel caching optimization can accelerate operations of FHE-based outsourced databases~\cite{tawose2023toward}. BatchCrypt~\cite{zhang2020batchcrypt} offers a system solution for distributed machine learning based on FHE, accelerating communication and encryption through data batching. 

\del{There are some attempts for the privacy problem in causal structure learning under distributed settings.
The work~\cite{ng2022towards} utilized continuous optimization to handle distributed structure learning problems, solving models among agents holding horizontally partitioned data. However, this method does not consider privacy protection and mainly aims at the theoretical optimization of the learning process. More recently, a privacy-aware method~\cite{huang2022towards} used a voting mechanism to decide the existence of edges for learning the causal structure, but the method is inaccurate in statistical tests and is an attempt at privacy-aware causal structure learning. Nevertheless, transmitting raw voting data also exposes private information to attackers having reserve deduction ability. In this paper, we aim to have full privacy protection while having accurate statistical results.}

Efforts have been made to address the privacy issue in distributed causal structure learning. The study by Ng et al.\cite{ng2022towards} employed continuous optimization to tackle distributed structure learning problems among agents holding horizontally partitioned data, though without considering privacy protection. A recent privacy-aware method\cite{huang2022towards} utilized a voting mechanism for causal structure learning, but it presents limitations in statistical test accuracy and exposes private voting data to potential threats. This paper strives to ensure comprehensive privacy protection while maintaining statistical accuracy.

\section{Conclusion}
\label{paper:con}
In this paper, we have proposed a privacy-preserving causal structure learning algorithm with fully homomorphic encryption (FHE).
The key challenges of this problem include complex arithmetic circuits, the requirement for division and logarithmic operations, and inefficiencies in execution. 
We have proposed a series of techniques to address these challenges including equation simplification, high-quality approximation for division and logarithm, and batching mechanisms to improve efficiency. A thorough examination of the experimental results reveals the efficacy of the proposed method. It is capable of generating causal structures of similar quality to those obtained from plaintext methods, demonstrating its effectiveness. Additionally, the method is not only efficient to deploy, but also exhibits commendable scalability and be seamlessly extended to support other privacy-preservation techniques (e.g., DP). 
\bibliography{myref}
\bibliographystyle{IEEEtran}

\end{document}